\documentclass[%
reprint,
superscriptaddress,
 amsmath,amssymb,
 aps,
 prb,
]{revtex4-2}

\usepackage{graphicx}
\usepackage{dcolumn}
\usepackage{bm}
\usepackage[normalem]{ulem}
\usepackage{xcolor}
\usepackage{mathrsfs}  

\renewcommand{\t}[1]{{\text{#1}}}

\begin{document}

\title{A frequency tunable low-noise YIG-GGG based oscillator \\with strong magneto-elastic coupling} 

\author{P. Sgarro} \thanks{Correspondence to: paolo.sgarro@cea.fr}
\affiliation{Department of Physics, University of Gothenburg, Gothenburg 41296, Sweden}
\affiliation{SPINTEC, Univ. Grenoble Alpes, CEA, CNRS, IRIG-SPINTEC, F-38000 Grenoble, France.}

\author{R. Ovcharov}
\affiliation{Department of Physics, University of Gothenburg, Gothenburg 41296, Sweden}

\author{R. Khymyn}
\affiliation{Department of Physics, University of Gothenburg, Gothenburg 41296, Sweden}

\author{S. Ghosh}
\affiliation{Department of Physics, University of Gothenburg, Gothenburg 41296, Sweden}

\author{A. A. Awad}
\affiliation{Department of Physics, University of Gothenburg, Gothenburg 41296, Sweden}
\affiliation{Center for Science and Innovation in Spintronics and Research Institute of Electrical Communication (CSIS), Tohoku University, 2-1-1 Katahira, Aoba-ku, Sendai 980-8577 Japan}
\affiliation{Research Institute of Electrical Communication (RIEC), Tohoku University, 2-1-1 Katahira, Aoba-ku, Sendai 980-8577 Japan}

\author{J. Åkerman}
\affiliation{Department of Physics, University of Gothenburg, Gothenburg 41296, Sweden}
\affiliation{Center for Science and Innovation in Spintronics and Research Institute of Electrical Communication (CSIS), Tohoku University, 2-1-1 Katahira, Aoba-ku, Sendai 980-8577 Japan}
\affiliation{Research Institute of Electrical Communication (RIEC), Tohoku University, 2-1-1 Katahira, Aoba-ku, Sendai 980-8577 Japan}

\author{A. Litvinenko} \thanks{Correspondence to: artem.litvinenko@physics.gu.se}
\affiliation{Department of Physics, University of Gothenburg, Gothenburg 41296, Sweden}


\date{\today}

\begin{abstract}

We present a frequency tunable magneto-acoustic oscillator (MAO) operating in low-phase-noise and complex dynamical regimes based on a single composite YIG-GGG resonator. The magneto-acoustic resonator (MAR) is based on a YIG (yttrium iron garnet) layer epitaxially grown on a GGG (gadolinium gallium garnet) substrate. By optimizing the YIG thickness, we obtain a high magneto-elastic coupling of around 1 MHz between the ferromagnetic resonance (FMR) in YIG and high overtone acoustic resonances (HBARs) in the YIG-GGG structure in the 1-2 GHz frequency range. It allows to eliminate the need for pre-selectors and bulky circulators, thus simplifying the MAO design while maintaining the possibility to lock to HBAR YIG-GGG modes. With an adjustment in the loop over-amplification parameter, the MAO can be locked either only to high-Q magneto-acoustic HBARs or to both types of resonance including HBARs and the FMR mode of the YIG film. In a low-phase-noise regime, MAO generates only at certain values of the applied field and exhibits discrete frequency tunability with a 3.281 MHz step corresponding to the frequency separation between the adjacent HBAR modes in a YIG-GGG structure. In a complex regime where oscillation conditions expand to include both HBAR and FMR modes, MAO demonstrates continuous generation as the function of the applied field with variable phase noise parameters. Moreover, in low-phase-noise regime, MAO phase noise plot improves by 30 dB compared to the operational regime locked to the pure FMR in YIG which is in agreement with the measured FMR and HBAR Q-factors.



\end{abstract}

\maketitle


\section{\label{sec:level1}Introduction}
The need for discretely \cite{vaucher2006architectures,stegner2018multi,chenakin2017frequency} and continuously \cite{park1999low,van2019low} frequency-tuned oscillators is fundamental to numerous signal processing applications in automotive, medical and communications systems. Depending on the purpose of the oscillator, there are different requirement for the output signals properties. For chaotic communication \cite{volkovskii2005spread}, random number generation \cite{phan2024unbiased}, and radio-frequency illumination \cite{dmitriev2016radio,dmitriev2017radio} application, it is required to have broad spectrum with high phase noise. However, for the most general applications, the phase noise of the oscillator should be as low a possible. 

The phase noise of an oscillator is mainly determined by the quality factor of its resonator. 
Several technologies, with different advantages and drawbacks, are offering high-Q resonators for high-frequency oscillator designs: dielectric resonators operate in the 0.1-100 GHz frequency range, and their quality factor reaches $10^3-10^4$, however, they are rather bulky; surface acoustic wave (SAW) resonators are usually used from 100 MHz to 3 GHz and also provide Q-factor of $10^3-10^4$, bulk acoustic wave (BAW) resonators \cite{liu2020materials,zhao2022x} including high-overtone BAW resonators (HBAR) \cite{Valle2021HBARs} provide with a similar q-factor but can operate up to 20 GHz, micro and nano-electro-mechanical system resonators (MEMS/NEMS) \cite{aubin2003laser, stassi2021reaching} operate at MHz frequency range and but have micro- and nano-scale size and Q-factors around $10^4$. While the described above solid-state resonators provide a high-Q factor and operate at a high frequency, they are essentially non-tunable, which limits their possible application range. Moreover, despite advancements and active development of frequency synthesis techniques \cite{vaucher2006architectures,chenakin2017frequency} that allow to exploit fixed-frequency resonators and mixed-signal processing blocks to generate arbitrary frequency, the need for resonators whose frequency can be rapidly tuned in wide range of frequencies still exists. 

\begin{figure*}[ht!]
    \centering
    \includegraphics[width=1.0\textwidth]{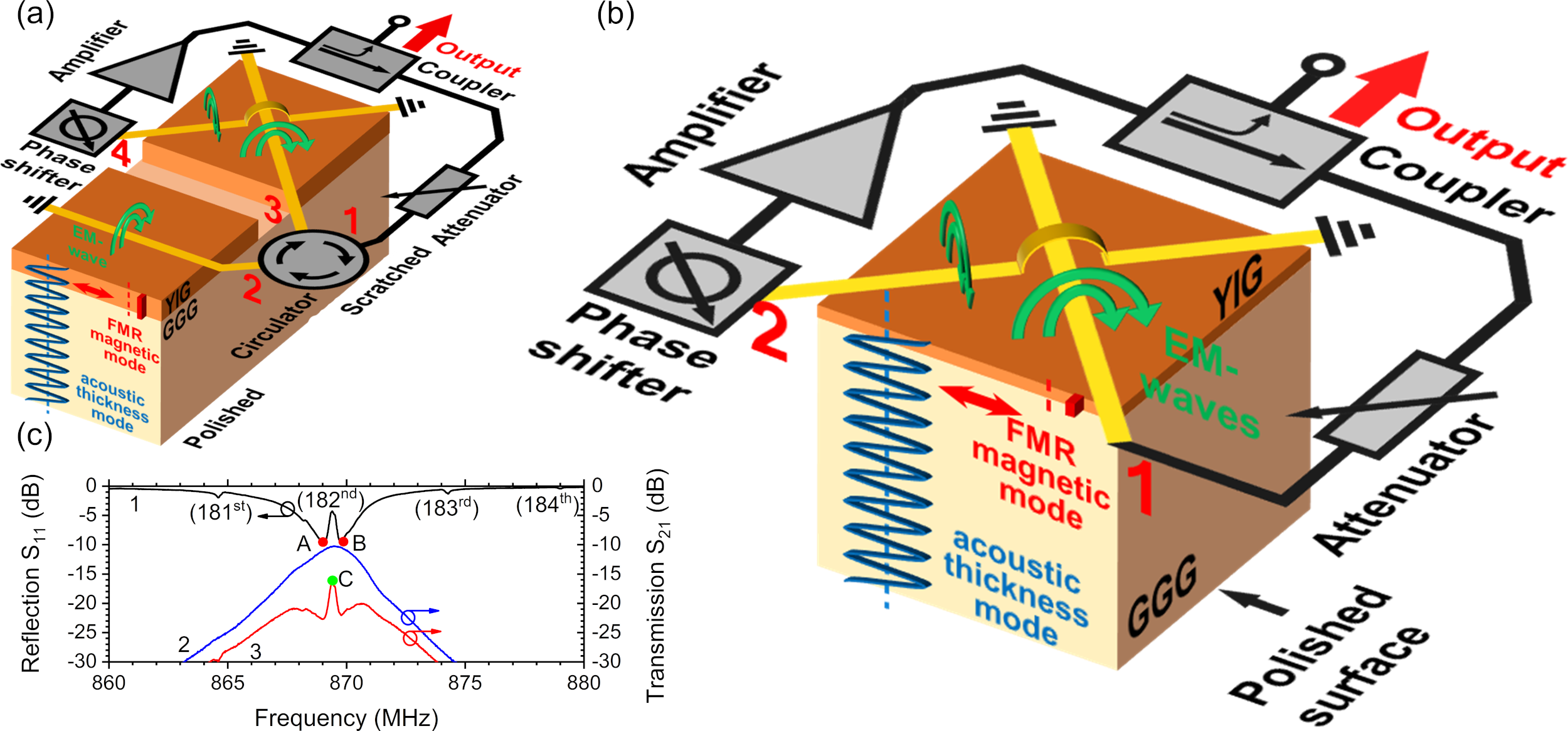}
    \caption{Schematics of YIG-GGG magneto-acoustic oscillators. (a) A previous MAO design presented in \cite{litvinenko2021tunable} comprises a one-port magneto-acoustical resonator, a two-port FMR preselector filter, and an external circulator. YIG thickness is $9.75\mu \t{m}$ (b) An improved MAO design with a two-port magneto-acoustic resonator with increased magneto-acoustic coupling due to the decreased YIG thickness of is $1.08\mu \t{m}$. Output and input microwire transducers are arranged with 90 degrees relative rotation to minimize inductive cross-coupling. Panel (c) displays the S-parameters ($S_{11}$ and $S_{21}$) in a previous design. Points A and B correspond to the possible locking frequencies in the one-port oscillator design and are away from the acoustic resonance of the 182nd YIG-GGG HBAR mode. The blue curve corresponds to the $S_{21}$ of the YIG FMR preselector filter. The red curve represents the $S_{21}$ parameter of serially connected resonators. Point C shows a stable locking frequency in a complex multi-resonator MAO design.}
    \label{fig:figure1}
\end{figure*}

One of the directions in frequency tunable oscillators is the use of magnetic YIG resonators and delay lines. \cite{Bankowski2015magnon,dai2020octave,gevorkyan2021ferrite,nikitin2024microwave}. The ferromagnetic resonance of the YIG layer is tunable since it depends on the strength of a magnetic field applied through an integrated permanent magnet. Conversely, damping losses in ferromagnet tend to be large (around 1 \%) and to increase with frequency \cite{serga2010yig}. Hence, the Q-factor of magnetic resonators is rather low as compared to acoustic and MEMS/NEMS resonators, especially at around 1 GHz \cite{lakin1993high}.

Previously, we demonstrated a high-Q magneto-acoustic resonator (MAR) \cite{litvinenko2015brillouin, tikhonov2016brillouin, litvinenko2021tunable} based on YIG/GGG composite layer that combines the advantages of frequency tunability from magnetic subsystem and high Q-factor of GGG acoustic subsystem by exploiting the large magnetostriction observed in YIG \cite{comstock1965magnetoelastic}, the lattice matching between the two garnets structure YIG and GGG, and their almost equal sound velocities \cite{gulyaev1981observation} \cite{kazakov1983magneto} \cite{zilberman1985self} \cite{gulyaev1988magnetoelastic}. In that design \cite{litvinenko2021tunable}, the YIG acts as a frequency-tunable transducer, which can selectively excite one of the standing acoustic modes of the YIG/GGG structure. However, since the magneto-elastic coupling was low due the combination of the frequency range used and particular thickness of the YIG film, the amplitude of acoustic peaks in $S_{11}$-parameter curve were rather small (see peaks around modes 181st, 183rd and 184th in Fig.~\ref{fig:figure1}(c)) which forced to exploit only central symmetrical acoustic peaks (mode 182nd in Fig.~\ref{fig:figure1}(c)). It required to convert one port MAR (black curve in Fig.~\ref{fig:figure1}(c)) into a two-port device with a standard microwave circulator and an additional YIG-preselector filter (blue curve in Fig.~\ref{fig:figure1}(c))to isolate conditions for oscillations around a central symmetric acoustic peak (red curve in Fig.~\ref{fig:figure1}(c)). While we successfully managed to implement a MAR-based oscillator (MAO) scheme that demonstrates a phase noise improvement of 20 dB as compared to a ferromagnetic resonance(FMR)-based oscillator, the use of an additional external microwave circulator with the lateral size of 10 cm and additional mechanical scratching of the YIG preselector filter made the technology of the MAO very complex. Here, we aim to improve the MAO design by enhancing magneto-elastic coupling, which allows for a classical two-port oscillator scheme to be implemented. The suggested two-port magneto-acoustic oscillator demonstrates two distinct oscillatory regimes that expands the range of potential application. In the first regime MAO, demonstrates discrete frequency tunability that locks to desired acoustic modes demonstrating low phase noise operation. The second regime allows MAO to switch between FMR mode and HBAR modes demonstrating phase noise improvement up to 30 dB at HBAR modes as compared to the regime when it is locked to the FMR mode. 

\begin{figure*}[ht!]
    \centering
    \includegraphics[width=\textwidth]{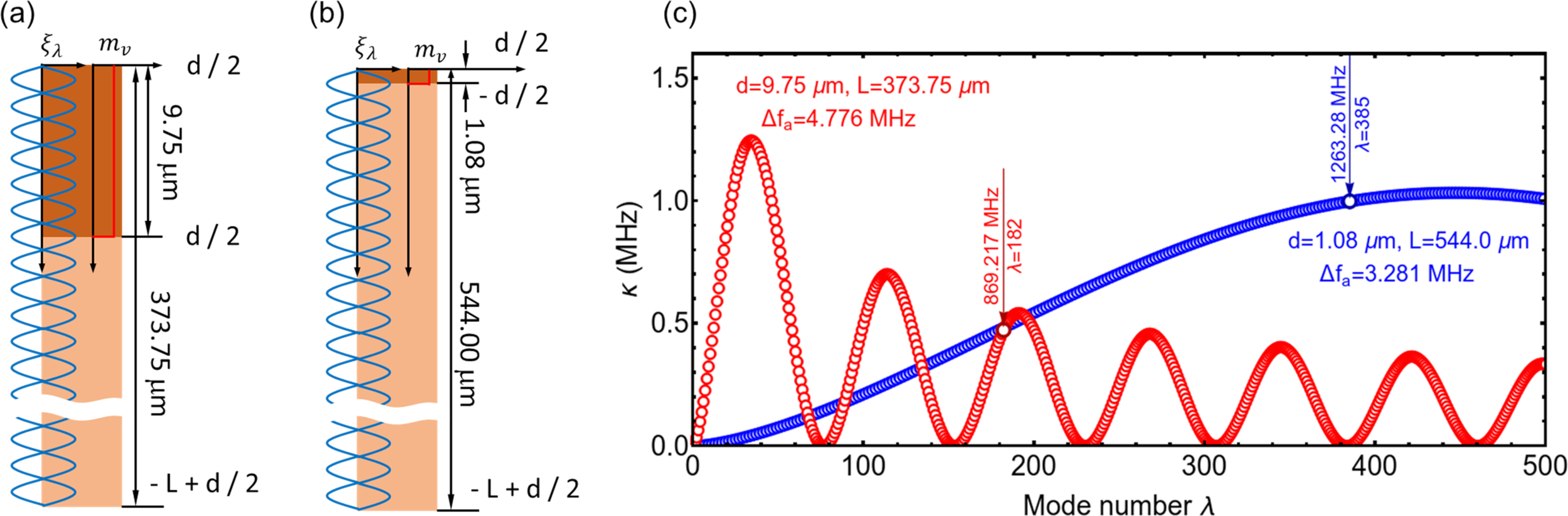}
    \caption{The thickness profiles of the magnetic FMR mode (red) and the standing acoustic mode (blue) in the YIG-GGG bilayer shown schematically for (a) the geometry studied in Ref.~\cite{litvinenko2021tunable} and (b) the geometry used in this study. Panel (c) shows the theoretically calculated magnetoacoustic coupling efficiency as a function of mode number, with red and blue colors indicating geometries (a) and (b), respectively.}
    \label{fig:figure2}
\end{figure*}

\section{Thickness dependence of the magnetoacoustic coupling}

The S-parameter characteristics of the MAR are highly dependent on the strength of the magnetoelastic coupling. This coupling can be calculated as a function of both frequency and geometry, based on the theoretical description of magnetoacoustic interactions within the structure. Considering the coupled equations for the ferromagnetic mode $\tilde{m}$ (with a uniform thickness profile $\tilde{m}=1$ for the FMR mode) and the $\lambda$-th acoustic mode (described by $\cos [(z - d/2) \pi \lambda / L]$) in the sample, the coupling constant $\kappa$ (measured in Hz) is given by~\cite{verba2018nonreciprocal, lisenkov2019magnetoelastic,brataas2020spin, litvinenko2021tunable}:
\begin{equation}\label{eq.kappa}
    \kappa^2 = \frac{\gamma b^2 }{8 \pi^3 M_s \lambda  d  \sqrt{c_{44} \rho}}  \left( 1 - \cos \frac{\pi \lambda d}{L} \right)^2
\end{equation}

where $\gamma/(2\pi)=28.3$~GHz/T is a gyromagnetic ratio of YIG, $b=7\times10^5$~J/m$^3$ is constant of magnetoelastic interaction, $M_s=138.5$~kA/m is saturation magnetization of YIG, $c_{44}=6.59\times 10^{10}$~N/m$^2$ is the elastic modulus of GGG, $\rho=5.17\times10^3$~kg/m$^3$ is mass density, $d$ is the thickness of the YIG film, and $L$ is the thickness of the bilayer.

In Fig.~\ref{fig:figure2}(c), Eq.~(\ref{eq.kappa}) is plotted against the mode number $\lambda$ for different geometries shown in Fig.~\ref{fig:figure2}(a,b). As can be seen, scaling down the thickness of the YIG magnetic subsystem improves the magnetoelastic coupling $\kappa$ from 0.47~MHz to 1.0~MHz which is significant and may lead to asymmetric acoustic resonance on the $S_{21}$-parameter of the resonator characteristic. In the previous design, $\kappa$ had a maximum at around the 50th and 110th modes. However, the corresponding frequencies for these modes 164.05~MHz and 360.91~MHz are too low for the efficient excitation of FMR. Hence, only the peaks around mode numbers 170th to 200th could be exploited. In the improved MAR geometry (see Fig.~\ref{fig:figure1}(b,c)), the coupling constant $\kappa$ is comparable with the previous design at mode numbers around 180th and keeps raising up to the optimal modes 300-600 with strong magneto-elastic coupling have frequencies 1-2~GHz at which the acoustic losses in GGG are still acceptable. While further reduction of the YIG thickness improves the absolute value of the magneto-acoustic coupling in the system, it moves the frequency range for optimal modes to 2-3~GHz, where the acoustic losses are much higher, and the Q-factor of the GGG acoustic subsystem is lower. Henceforth, for practical oscillator design scaling the YIG thickness below 500-800~nm has no impact.

\section{Two-port MAR}

The device developed is a parallel-plate straight-edge rectangle YIG-GGG resonator cut from a monocrystalline epitaxial heterostructure. The YIG thickness is 1.08 $\mu m$ and the GGG thickness is 543 $\mu m$; the magnetization is saturated in-plane that allows to significantly reduce the absolute value of the required applied magnetic field from previously reported 215 mT \cite{litvinenko2021tunable} to mere 10.4 mT. Two microstrip transducers are used to get a two-port device and are rotated relative to one another by 90 degrees to avoid inductive cross-coupling (see Fig.~\ref{fig:figure1}(b)) The design alone constitutes an evident improvement with respect to the previous work, enabling to remove the two-port preselector and the bulky circulator.

\begin{figure*}[ht!]
    \centering
    \includegraphics[width=\textwidth]{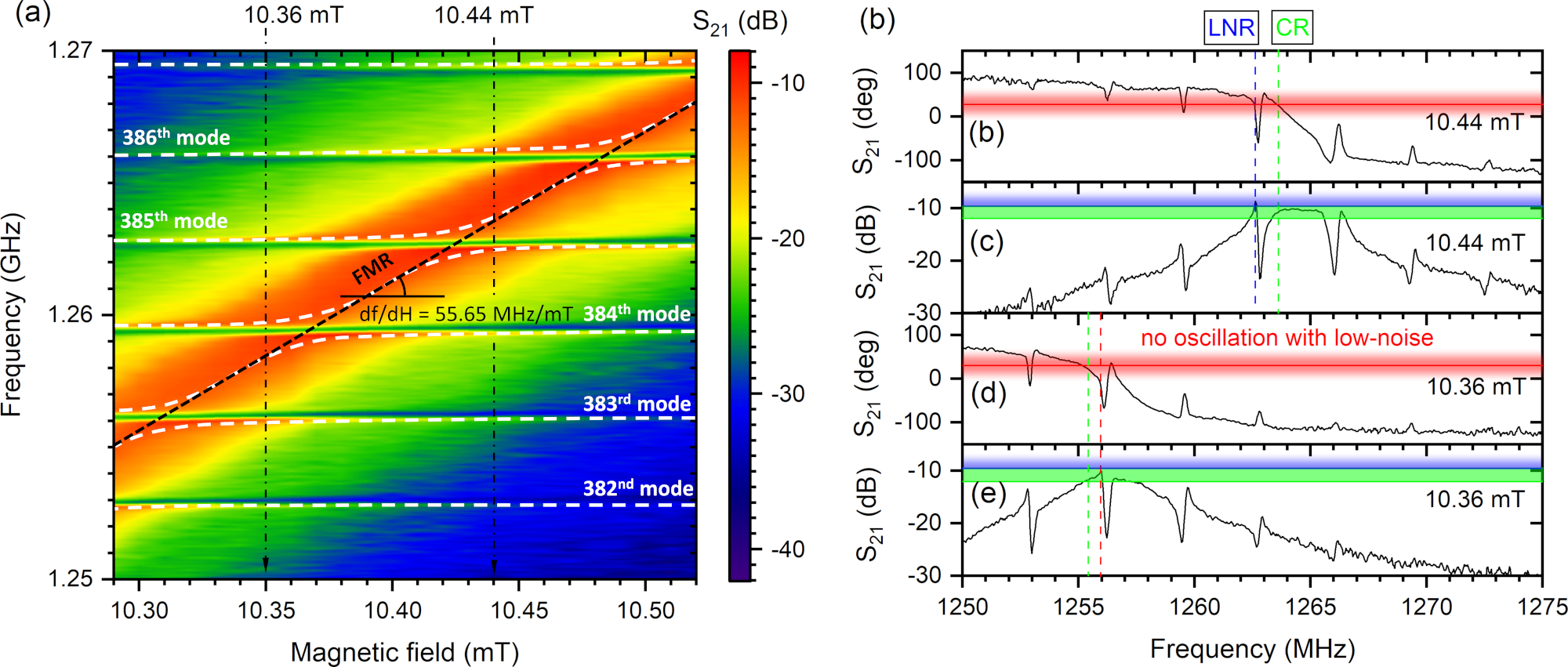}
    \caption{Characterization of the $S_{21}$-parameter between 1.25 and 1.27 GHz, with an applied magnetic field between 10.29 and 10.52 MHz. The full spectrogram (a) shows the pronounced avoided crossings between the magnetic modes and the HBARs; and the frequency tunability; i.e., the FMR frequency dependence with the applied field. The phase (c) (e) and the amplitude (b) (d) of the $S_{21}$-parameter show a large asymmetry, a sign of strong magnetoacoustic coupling, and can be used to identify the stability of auto-oscillations, as a consequence of Barkhausen criterion, and to distinguish between coupled regime, and low-noise regime.}
    \label{fig:figure3}
\end{figure*}

First, we characterize the magnitude of $S_{21}$-parameter as the function of frequency for different applied magnetic fields, which provides the details on the tunability of the device. As can be seen from Fig.~\ref{fig:figure3}(a), the FMR frequency grows with the field with a slope of $df/dH = 55.65$ MHz/mT which is close to the theoretical value, which can be obtained by taking the linear term of the FMR frequency Taylor series expansion
\begin{equation}
f_{\text{FMR}}=\gamma\sqrt{H(H+M_s)},
\end{equation}
\begin{equation}
df/dH=\gamma\frac{2H_0+M_s}{2\sqrt{H_0(H_0+M_s)}}=62.84~\text{MHz/mT},
\end{equation}

We note that in contrast to our previous work~\cite{litvinenko2021tunable}, where the frequency tunability slope was 28.3~MHz/mT, here it is enhanced by a factor of two due to the use of in-plane magnetization geometry. We also note that the total transmission losses in the $S_{21}$-parameter at the FMR and acoustic resonances are relatively high, which can be explained by an impedance mismatch between electromagnetic wave excited by a narrow but long wire connected to a 50 Ohm cable and FMR standing spinwave mode. In~\cite{van2019low}, it was demonstrated with YIG spheres that transmission losses in two-port resonator based on orthogonal bond wire loops and FMR mode of YIG resonator can be as low as 6 dB even at 20 GHz. We believe that impedance mismatching in MAR can also be significantly improved with the right combination of sample size, length, and width of the coupling orthogonal wires. 

A strong magneto-elastic coupling in MAR leads to very well-pronounced avoided mode crossings \cite{An2020, an2022bright} between the magnetic mode and high-overtone bulk acoustic resonances (HBAR) (see Fig.~\ref{fig:figure3}(a)). The main feature of the individual $S_{21}$-parameter characteristics of a multiphysical MAR is the asymmetric shape of acoustic resonances when they are detuned from the FMR central frequency (see both Fig.~\ref{fig:figure1}(c) and Fig.~\ref{fig:figure3}(c,e)). Such asymmetric behavior is always present in coupled multi-physical systems but becomes more pronounced at strong coupling coefficients. For example, at the field 10.44 mT (see Fig.~\ref{fig:figure3}(c)), the magnitude of $S_{21}$-parameter exceeds -10 dB attenuation only in a very narrow frequency range around acoustic mode  $385^{th}$ while its phase changes significantly (see Fig.~\ref{fig:figure3}(b)). On the contrary, when FMR coincides with the central frequency of a certain acoustic mode, for example, at the field 10.36 mT, the acoustic resonance appears as a symmetric deep on the background of a wide FMR peak.

The main characteristic of a resonator is its quality factor (Q-factor). Using the Darko Kajfez method, we successfully extracted Q-factors for both HBAR and YIG FMR subsystems despite the unconventional shape of magneto-acoustic resonances. The details can be found in Appendix 1. According to our calculation, HBAR demonstrates a Q-factor of 9018, while FMR shows only 446. This allows us to expect a significant improvement in the phase noise figure if MAO switches locking from FMR mode to HBAR resonances of the YIG-GGG structure.  

\section{MAO circuit design}

\begin{figure*}[ht!]
    \centering
    \includegraphics[width=\textwidth]{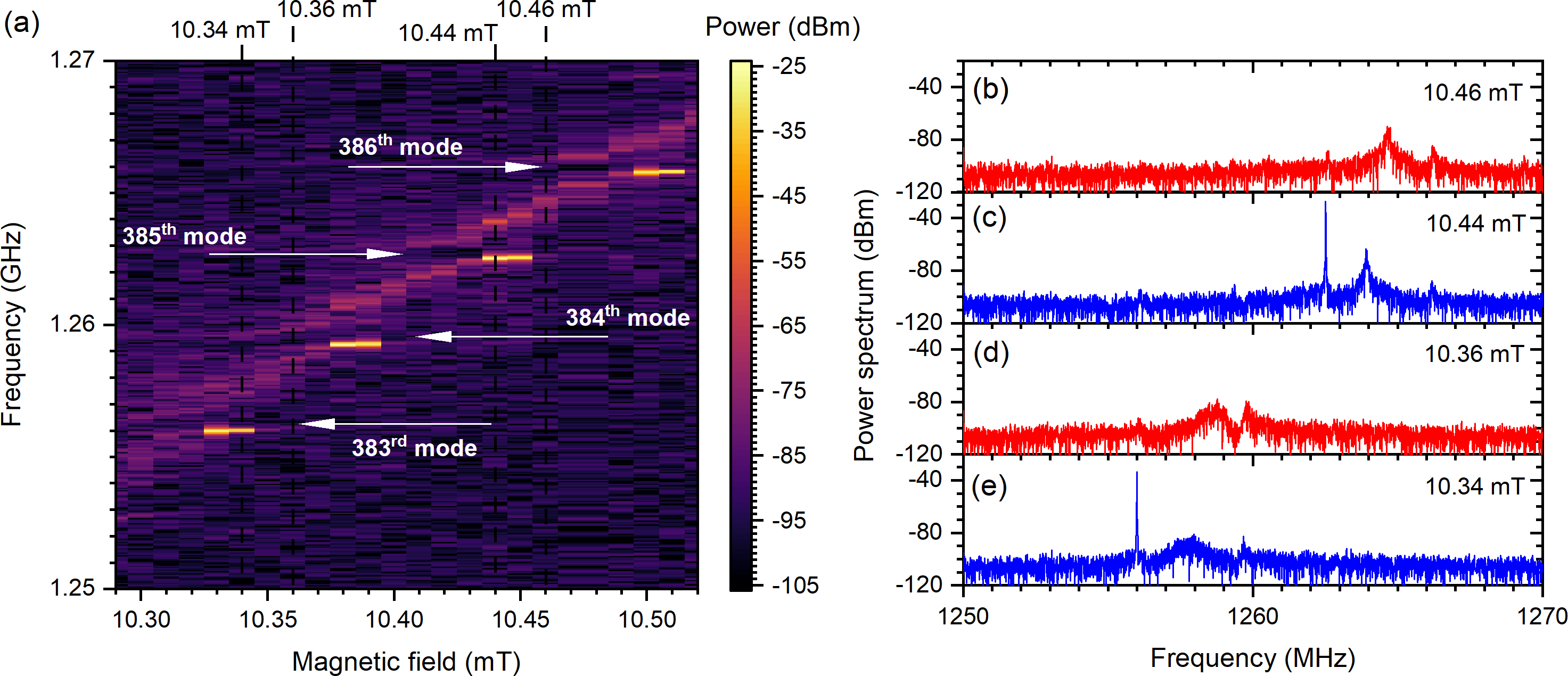}
    \caption{Characterization of MAO output signal. The amplitude of the spectrogram and individual spectra is corrected by a coupling coefficient of 12 dB due to the output coupler. (a) Spectrogram of the output signal as a function of applied magnetic field. (b) MAO individual output spectrum at 10.46 mT for pre-oscillatory dynamics. The conditions for auto-oscillation are not satisfied. (c) MAO individual output spectrum at 10.44 mT. A small linewidth signal is locked to the 385th acoustic mode of the YIG-GGG structure.  (d) MAO individual output spectrum at 10.36 mT for pre-oscillatory dynamics. The conditions for auto-oscillation are not satisfied. (e) MAO individual output spectrum at 10.34 mT. The low linewidth signal is locked to the 383rd acoustic mode of the YIG-GGG structure.}
    \label{fig:figure4}
\end{figure*}

The asymmetric shape of acoustic resonances, together with a large deviation of phase due to strong coupling, opens up the potential to design a two-port oscillator scheme to lock onto high-Q acoustic resonances of the YIG/GGG structure as both Barkhausen criteria for stability of oscillations can get satisfied:
\begin{align}
        \beta \mathbf{A} &= 1, \label{eq:BarkhausenStabAmpl} \\
        \angle \beta \mathbf{A} &= 2\pi n,\: n \in \{0,1,2...\} 
    \label{eq:BarkhausenStabPhase}
\end{align}
where $\mathbf{A}$ is the total amplification in the loop, $\beta$ is the total loss, and $\angle \beta \mathbf{A}$ is the phase accumulated along the main loop. 

In order to satisfy the condition (\ref{eq:BarkhausenStabAmpl}), we introduced in the ring a linear amplifier Mini-circuits ZKL-33ULN-S+ with 30 dB of gain. Note that we used ready-from-the-shelf components having a fixed gain parameter that would significantly overcompensate the losses in the MAR, leading to nonlinear processes \cite{tikhonov2016brillouin} and possible chaotization \cite{litvinenko2018chaotic} if used directly before the MAR. Therefore, in order to avoid large over-amplification in the loop and possible nonlinear processes we added a variable attenuator tuned to 8 dB for the low-noise to the input of the MAR. Next, in order to satisfy the condition (\ref{eq:BarkhausenStabPhase}) the variable phase shifter was tuned so that the overall phase accumulation at the frequency with the maximum of $S_{21}$-parameter magnitude (depicted in with blue dashed vertical line in Fig.~\ref{fig:figure3}(b,c)) in the MAR as well as in a phase shifter, coupler, amplifiers, and cabled is integer proportional to $2\pi$. In Fig.~\ref{fig:figure3}(c,e), the blue horizontal line depicts the minimum of $S_{21}$-parameter magnitude, which satisfies the amplitude Barkhausen criterion (\ref{eq:BarkhausenStabAmpl}). The blue line corresponds to the case when the variable attenuator is set to 8 dB. In Fig.~\ref{fig:figure3}(c,e), the red solid line corresponds to the value of $S_{21}$-parameter phase, which satisfies the phase Barkhausen criterion (\ref{eq:BarkhausenStabPhase}). We highlighted the area around +-30deg in phase since the Barkhausen criterion for phase can auto-adjust due to the nonlinearity of the amplifier and the MAR still allowing for oscillations. Nevertheless, as can be seen from Fig.~\ref{fig:figure3}(b-e), MAO may only operate at a frequency very close to the 385th acoustic mode, and an applied magnetization field is 10.44 mT. In the case of the applied field 10.36 mT there are no frequencies at which the Barkhausen criteria for oscillations get satisfied. Thus, when the variable attenuator is tuned to 8 dB, we can expect only a low-phase noise regime when acoustic mode appears as asymmetric peaks on the side of the FMR. 

\section{Results on operational regimes and phase noise characterization}

\begin{figure*}[ht!]
    \centering
    \includegraphics[width=1.0\textwidth]{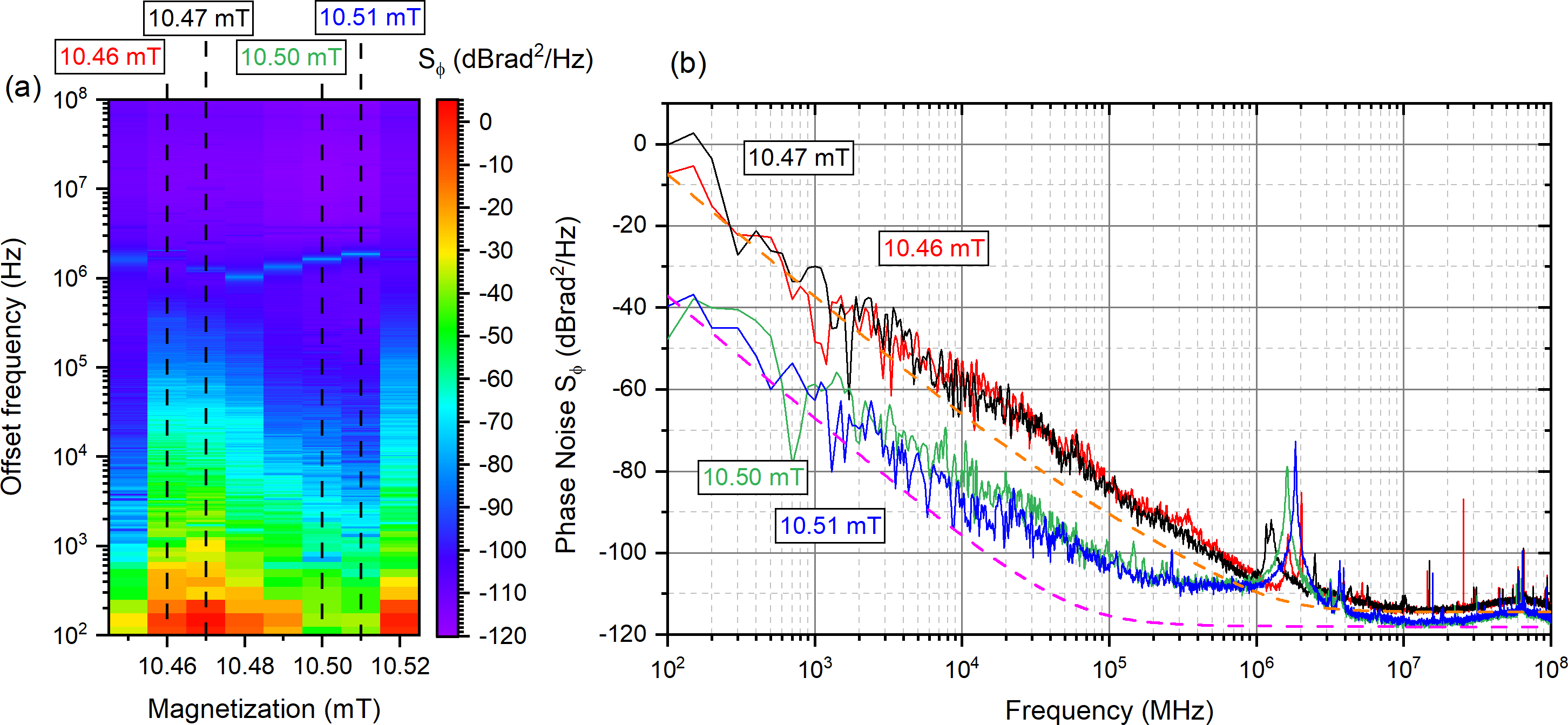}
    \caption{Phase noise characterization of the MAR. (a) The phase noise spectrogram as a function of the applied magnetic field. The results of sections at field values 10.46, 10.47, 10.50, and 10.51 mT are presented in the form of individual phase noise plots in panel (b). The phase noise plot at 10.46 and 10.47 mT demonstrate values around - 35 dBrad$^2$/Hz at 1kHz offset frequency due to the locking to the FMR mode. The phase noise significantly improves at 10.50 and 10.51 mT reaching - 65 dBrad$^2$/Hz at 1kHz offset frequency due to the locking to the $386^{th}$ HBAR mode. The orange and magenta dashed lines correspond to the results obtained with empirical Leeson's law taking into account the output power of -26 dBm and power at the output of the resonator of -51 dBm.}
    \label{fig:figure5}
\end{figure*}

We have demonstrated that the Barkhausen criteria for oscillations can be isolated around acoustic peaks in a simplified ring-type MAO. The next step is to study the properties of the oscillator's signal. We demonstrate two regimes that differ by a value of additional attenuation. The first regime, when the attenuator is set to 8 dB, allows only a low-phase regime in MAO. In the figure Fig.~\ref{fig:figure4}(a), we present a spectrogram of the MAO as a function of applied field in the same range as in Fig.~\ref{fig:figure3}(a). As can be seen, in this configuration MAO oscillates only at certain values of the applied field and exhibits a single frequency regime without parasitic self-modulation, chaotization \cite{litvinenko2018chaotic} or formation of dense trains of spin-wave solitons as could be observed in large-delay YIG waveguides \cite{ustinov2009generation, ustinov2014self,bir2020experimental,bir2024direct}. Surprisingly, the output power of the MAO is only -26 dBm, while a ZKL-33ULN-S+ amplifier has a high 1dB compression point for its output signal of 17 dBm. It indicates that the amplifier operates in a linear regime, and the saturation of the oscillator's power happens at the MAR itself. We attribute this low nonlinear threshold to several factors such as a relatively small thickness \cite{massouras2024mode, srivastava2023YIGeigen} and area of the YIG film, in-plane magnetization configuration \cite{demokritov2001brillouin, an2024emergent} and strong magnetoelastic coupling \cite{tikhonov2016brillouin}. Noteworthy, there are still pre-oscillatory frequency components, for example, at the field 10.36 and 10.46 mT, that reveal avoided mode crossings shown with white lines in Fig.~\ref{fig:figure3}(a) and can be attributed to the FMR mode but their amplitude is rather low.

The oscillatory conditions can be expanded to enable a second complex regime when MAO demonstrates oscillations at any value of the applied field. In this case, MAO locks not only to acoustic modes but also to FMR mode at certain values of the applied field. For this regime, the variable attenuator should be set to 6 dB. It corresponds to a green horizontal line in Fig.~\ref{fig:figure4}(b,c). The results of the signal analysis are presented in the form of the phase noise spectrogram (see Fig.~\ref{fig:figure5}(a)) and individual phase noise characteristics (see Fig.~\ref{fig:figure5}(b)). Note that here we present the phase noise in the $S_{\phi}(f)$ form, which is more appropriate \cite{rubiola2006measurement, rubiola2008phase} for high-precision measurements and scientific work as compared to archaic $\mathcal{L}(f)$. As expected in the case of ring oscillator architecture with a single loop, MAO demonstrates single-frequency operation at any value of the applied field. As expected, the corresponding phase noise changes significantly. At the fields 10.46 and 10.47 mT, the MAO locks to the FMR, which defines its phase noise at the level of around -35 dBrad$^2$/Hz at 1kHz offset frequency, while at the fields 10.51 and 10.50 mT MAO demonstrates the level of around -65 dBrad$^2$/Hz. These results correspond well to the empirical estimations from Leeson's equation (\ref{eq:1}), which are shown as magenta and orange dashed lines in Fig.~\ref{fig:figure6}(b). We note that in the previously considered low noise regime, the phase noise also reaches -65 dBrad$^2$/Hz at 1kHz offset frequency. Thus, this comparison shows a significant improvement in the low-phase noise, up to 30 dBrad$^2$/Hz, with respect to the case when MAO locks to the FMR mode. This noise reduction is also significantly larger than the one obtained in the 4-ports MAR, which was only 20 dB better than the magnetic one.

The presented oscillation regimes are achieved by careful tuning of the total amplification value in the loop, large enough to enable the acoustic or FMR modes but small enough to ensure a single-frequency oscillation regime, leading to theoretically predicted levels of the phase noise according to the Q-factor of the acting resonators and signal power levels at the output of the MAR.

\section{Conclusions and outlook}
We developed and characterized a tunable single-resonator magneto-acoustic oscillator (MAO) based on a YIG-GGG substrate with high magneto-elastic coupling between the ferromagnetic resonance (FMR) in YIG and the high overtone bulk acoustic resonances (HBARs) in the YIG-GGG structure. The optimization of the YIG thickness allowed us to enhance the coupling strength in the 1-2 GHz frequency range and simplify the oscillator design by removing an FMR-based YIG preselector and bulky circulator. The redesigned MAO can lock to either the FMR mode or to high-Q magneto-acoustic HBARs, depending on the overall loop gain.

We demonstrated that the MAO can operate in two distinct regimes. In the low-phase noise regime, the oscillator can only lock to the high-Q acoustic HBAR modes at certain values of the applied field, providing discrete frequency tunability with a 3.281 MHz step. This regime can be exploited for frequency synthesizers. In the complex regime, the oscillator is continuously tunable in frequency with the applied magnetic field, smoothly changing from the FMR-dominated to HBAR-dominated modes and demonstrating a phase noise improvement of up to 30 dB compared to the FMR-based regime.

Finally, in the presented MAO, the required external field is reduced from a previous 220mT to a mere 10 mT due to the use of in-plane magnetization geometry. Furthermore, the frequency tuning is enhanced more than twice, reaching 55.65 MHz/mT. Altogether, these parameters lead to a more practical design.

As an outlook, we foresee the possibility to sufficiently increase an output power and improving the absolute values of the phase noise in the MAO by increasing nonlinear saturation thresholds in MAR with improved impedance matching in microwave-to-spinwave transducers~\cite{devitt2024edge}, lateral sample size, and overall microwave design. These studies will improve the absolute figure of merit of the proposed MAO in the context of real-life signal processing and communication systems. Moreover, there is a prospect to increase magneto-acoustical coupling even further and move to higher frequencies if short dipole-exchange and exchange~\cite{tikhonov2020exchange, tikhonov2023excitation} spinwaves are employed. It is theoretically predicted that shear acoustic wave can be effectively excited by propagating and standing exchange spinwaves generated in YIG ferromagnetic resonators with non-uniform magnetization distribution ~\cite{TikhonovPhotonMagPhon2024}.

\section{Acknowledgements}
This project has received funding from the European Union’s 2020 research and innovation programme under the Marie Sklodowska-Curie grant agreement No. 955671. This work was partly supported by the Horizon 2020 research and innovation programme, ERC Advanced Grant No. 835068 “TOPSPIN”, the ERC Proof of Concept Grant No. 101069424 “SPINTOP” and the Marie Skłodowska-Curie grant agreement No. 101111429 “SWIM”.

\appendix
\section{Device fabrication}

The structures of the one-port reflection-type MAR and the composite two-port MAR are shown, respectively, in Fig.~1 and Fig.~4, Both resonators are manufactured using a two-layer parallel-plate YIG/GGG structure with the YIG thickness of $d=9.75\mu \t{m}$, GGG thickness of $L=364\mu \t{m}$, and YIG saturation magnetization of $M_{0} = 1740/(4\pi)Oe$.  In order to ensure highly parallel surfaces of the YIG/GGG structure, we made use of a chemical-mechanical polishing technique, achieving a wedge angle of less than $2''$, which ensures the formation of high overtone acoustic thickness resonances in the YIG/GGG structure. The fabricated one-port MAR had lateral dimensions of 1x1mm. The composite two-port MAR was fabricated by means of laser scribing technique\cite{Sheshukova2014} and had lateral dimensions of 1x2.5mm. For the two-port MAR, a scratch at the inner surface was needed to eliminate acoustic resonances as they would interfere with those in the one-port MAR; for this purpose, a micro-mesh wet sanding abrasive paper having grit 3200 was employed.

\section{Device characterization}
The developed two-port MAR was characterized using a vector network analyzer (Keysight Agilent E5071B). The phase noise of the developed composite MAO was measured via spectrum analysis of the instantaneous phase signal obtained with a Hilbert transform method. This method, exploited in the studies \cite{litvinenko2021analog, litvinenko2023phase, bianchini2010},  allows to extraction of the instantaneous phase of a time trace by reconstructing a corresponding complex signal $x(t)$ from the measured signal $v(t)$, as follows:
\begin{equation}
    x(t) = v(t) + j \mathscr{HT} [v(t)] = A(t)e^{j\phi(t)}
\end{equation}
where $\mathscr{HT} [v(t)]$ is the Hilbert transform of $v(t)$, $A(t)$ is time-varying amplitude of complex modulus $x(t)$ and $\phi(t)$ is its instantaneous phase.

Practically, time domain signal $v(t)$ is captured with a 12-bit 5GS/s oscilloscope (Teledyne LeCroy WaveRunner 8208HD). Then, Hilbert transform is done in MATLAB with built-in function hilbert() to obtain $x(t)$ and the instantaneous phase $\phi(t)$ signal is extracted with unwrap(angle()) command over $x(t)$ signal. The last step is a discrete Fourier transform of $\phi(t)$ to visualize phase noise spectra $S_{\phi}$ having the dimension of dBrad$^2$/Hz. 

While the precision of this method strongly depends on the resolution of the oscilloscope, it allows obtaining single-sideband phase noise spectral density or power spectral density of the phase fluctuations $S_{\phi}(f)$, which is more appropriate for high-precision measurements and scientific work as compared to archaic $\mathcal{L}(f)$ \cite{rubiola2006measurement, rubiola2008phase}.

\begin{figure}[ht!]
    \centering
    \includegraphics[width=0.47\textwidth]{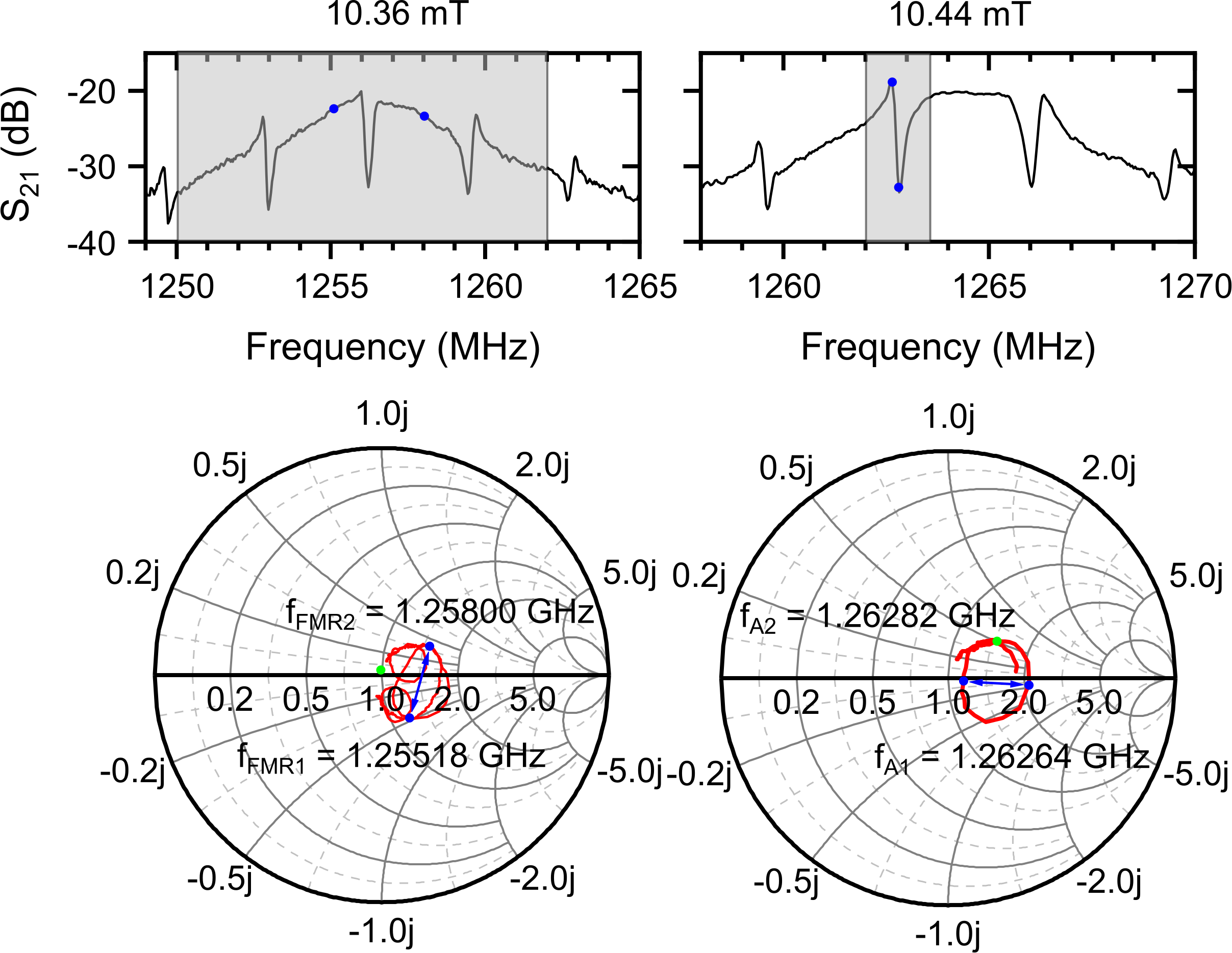}
    \caption{Frequency response and Smith chart representation of the MAR at different applied fields. Panels (a) and (b) show the transmission parameter $S_{21}$ at external magnetic fields of 10.36 mT and 10.44 mT, respectively. (a) An FMR mode with a quality factor $Q_{\text{FMR}} = 446$ is highlighted with a gray background. (b) A $385^{th}$ HBAR mode with an enhanced quality factor $Q_{A} = 9018$ is highlighted with a gray background. (c, d) Smith charts demonstrating resonance loops corresponding to the FMR and HBAR modes, correspondingly. The range of data points in Smith charts is limited, with frequency points inside the gray areas in the corresponding magnitude plots. The blue points are used to calculate the width of the resonances. The FMR edge frequencies are $f_{\text{FMR1}} = 1.25518$ GHz and $f_{\text{FMR2}} = 1.25800$ GHz. HBAR edge frequencies are $f_{A1} = 1.26264$ GHz and $f_{A2} = 1.26282$ GHz.
    }
    \label{fig:figure6}
\end{figure}

Characterization of the MAR Q-factors for acoustic and FMR resonances was done with a modified Darko Kajfez's method \cite{kajfez1986dielectric} with measured $S_{21}$-parameters plotted in a Smith Chart. This method is based on the analysis of the resonance loops in an $S$-parameter graph plotted on a Smith chart. This method allows us to deduce a dimensionless value of the loaded Q-factors. To obtain the $S_{21}$-parameter of the two-port MAR, we used VNA Keysight (Agilent E8361A). The wire antennas used in our experiments had a width of 0.2 mm and a length of 10 mm, which accounts for 9.12 nH and allows for a good matching between FMR in YIG and electromagnetic wave in 50-Ohm cables. In order to obtain the loaded Q-factors of the FMR and acoustic mode, a resonant loop corresponding to a particular resonance should be identified as shown in Fig.~\ref{fig:figure6}(a) for FMR mode and in Fig.~\ref{fig:figure6}(b) for 385th acoustic mode of YIG-GGG structure with highlighted gray areas. Then, the $S_{21}$-parameter from the highlighted areas is plotted in corresponding Smith Charts. The following geometric operations have to be done in order to get the Q-factor. First, the position of a point where the loop line crosses itself or is expected to cross itself has to be identified. In the case of the FMR loop, all the loops that correspond to acoustic resonances have to be ignored. Moreover, to observe the crossing a very wide frequency range has to be take and, therefore, it has to be estimated where it aims to cross. In the case of acoustic mode, the resonance loop has a relatively well-defined position of self-crossing, taking into account measurement noise. In both cases, we mark the crossing point with green points. The precision of the first operation is not that important as it just allows us to identify the orientation and the position of the diameter section auxiliary line that goes through the center of the loop. The blue points $f_{FMR1}$ and $f_{FMR2}$ in the case of FMR and $f_{A1}$ and $f_{A2}$ in the case of acoustic mode should be located at the intersection of the diameter auxiliary lines and the $S_{21}$-parameter curve and should be equidistant to the green points of loop self-crossing. The final step is to extract from the S-parameter file the frequency values that correspond to the identified blue points. 

The frequency gap $f_{FMR2}-f_{FMR1}$ defines the FMR linewidth and is used to calculate the loaded Q-factor of the FMR mode of the YIG film:
\begin{equation}
Q_{\t{FMR}}=\frac{f_{0}}{f_{FMR2}-f_{FMR1}},
\label{QfactorYIG}
\end{equation}

Similarly, the frequency interval $f_{A2}-f_{A1}$ defines the acoustic mode linewidth that is used to calculate the loaded Q-factor of the 385th acoustic mode of the composite YIG-GGG structure:
\begin{equation}
Q_{\t{MAR}}=\frac{f_{0}}{f_{21}-f_{11}},
\label{eq:QfactorMAR}
\end{equation}

We note that while an approach of measuring the linewidth at -3dB gives a similar Q-factor for FMR mode, it cannot be applied to asymmetric acoustic resonances of the YIG-GGG structure. According to (~\ref{eq:QfactorMAR}) and (~\ref{eq:QfactorMAR}), Q-factors of FMR mode and acoustic resonance of YIG-GGG are 446 and 9018, correspondingly. Detailed explanation and derivation of the above-presented approach and operations can be found in \cite{Kajfez1984}.

\bibliography{references.bib}

\begin{thebibliography}{54}%
\makeatletter
\providecommand \@ifxundefined [1]{%
 \@ifx{#1\undefined}
}%
\providecommand \@ifnum [1]{%
 \ifnum #1\expandafter \@firstoftwo
 \else \expandafter \@secondoftwo
 \fi
}%
\providecommand \@ifx [1]{%
 \ifx #1\expandafter \@firstoftwo
 \else \expandafter \@secondoftwo
 \fi
}%
\providecommand \natexlab [1]{#1}%
\providecommand \enquote  [1]{``#1''}%
\providecommand \bibnamefont  [1]{#1}%
\providecommand \bibfnamefont [1]{#1}%
\providecommand \citenamefont [1]{#1}%
\providecommand \href@noop [0]{\@secondoftwo}%
\providecommand \href [0]{\begingroup \@sanitize@url \@href}%
\providecommand \@href[1]{\@@startlink{#1}\@@href}%
\providecommand \@@href[1]{\endgroup#1\@@endlink}%
\providecommand \@sanitize@url [0]{\catcode `\\12\catcode `\$12\catcode `\&12\catcode `\#12\catcode `\^12\catcode `\_12\catcode `\%12\relax}%
\providecommand \@@startlink[1]{}%
\providecommand \@@endlink[0]{}%
\providecommand \url  [0]{\begingroup\@sanitize@url \@url }%
\providecommand \@url [1]{\endgroup\@href {#1}{\urlprefix }}%
\providecommand \urlprefix  [0]{URL }%
\providecommand \Eprint [0]{\href }%
\providecommand \doibase [0]{https://doi.org/}%
\providecommand \selectlanguage [0]{\@gobble}%
\providecommand \bibinfo  [0]{\@secondoftwo}%
\providecommand \bibfield  [0]{\@secondoftwo}%
\providecommand \translation [1]{[#1]}%
\providecommand \BibitemOpen [0]{}%
\providecommand \bibitemStop [0]{}%
\providecommand \bibitemNoStop [0]{.\EOS\space}%
\providecommand \EOS [0]{\spacefactor3000\relax}%
\providecommand \BibitemShut  [1]{\csname bibitem#1\endcsname}%
\let\auto@bib@innerbib\@empty
\bibitem [{\citenamefont {Vaucher}(2006)}]{vaucher2006architectures}%
  \BibitemOpen
  \bibfield  {author} {\bibinfo {author} {\bibfnamefont {C.~S.}\ \bibnamefont {Vaucher}},\ }\href@noop {} {\emph {\bibinfo {title} {Architectures for RF frequency synthesizers}}},\ Vol.\ \bibinfo {volume} {693}\ (\bibinfo  {publisher} {Springer Science \& Business Media},\ \bibinfo {year} {2006})\BibitemShut {NoStop}%
\bibitem [{\citenamefont {Stegner}\ \emph {et~al.}(2018)\citenamefont {Stegner}, \citenamefont {Fischer}, \citenamefont {Gropp}, \citenamefont {Stehr}, \citenamefont {M{\"u}ller}, \citenamefont {Hoffmann},\ and\ \citenamefont {Hein}}]{stegner2018multi}%
  \BibitemOpen
  \bibfield  {author} {\bibinfo {author} {\bibfnamefont {J.}~\bibnamefont {Stegner}}, \bibinfo {author} {\bibfnamefont {M.}~\bibnamefont {Fischer}}, \bibinfo {author} {\bibfnamefont {S.}~\bibnamefont {Gropp}}, \bibinfo {author} {\bibfnamefont {U.}~\bibnamefont {Stehr}}, \bibinfo {author} {\bibfnamefont {J.}~\bibnamefont {M{\"u}ller}}, \bibinfo {author} {\bibfnamefont {M.}~\bibnamefont {Hoffmann}},\ and\ \bibinfo {author} {\bibfnamefont {M.~A.}\ \bibnamefont {Hein}},\ }\bibfield  {title} {\bibinfo {title} {A multi-frequency mems-based rf oscillator covering the range from 11.7 mhz to 1.9 ghz},\ }in\ \href@noop {} {\emph {\bibinfo {booktitle} {2018 IEEE/MTT-S International Microwave Symposium-IMS}}}\ (\bibinfo {organization} {IEEE},\ \bibinfo {year} {2018})\ pp.\ \bibinfo {pages} {575--578}\BibitemShut {NoStop}%
\bibitem [{\citenamefont {Chenakin}(2017)}]{chenakin2017frequency}%
  \BibitemOpen
  \bibfield  {author} {\bibinfo {author} {\bibfnamefont {A.}~\bibnamefont {Chenakin}},\ }\bibfield  {title} {\bibinfo {title} {Frequency synthesis: Current status and future projections.},\ }\href@noop {} {\bibfield  {journal} {\bibinfo  {journal} {Microwave Journal}\ }\textbf {\bibinfo {volume} {60}} (\bibinfo {year} {2017})}\BibitemShut {NoStop}%
\bibitem [{\citenamefont {Park}\ and\ \citenamefont {Kim}(1999)}]{park1999low}%
  \BibitemOpen
  \bibfield  {author} {\bibinfo {author} {\bibfnamefont {C.-H.}\ \bibnamefont {Park}}\ and\ \bibinfo {author} {\bibfnamefont {B.}~\bibnamefont {Kim}},\ }\bibfield  {title} {\bibinfo {title} {A low-noise, 900-mhz vco in 0.6-/spl mu/m cmos},\ }\href@noop {} {\bibfield  {journal} {\bibinfo  {journal} {IEEE Journal of Solid-State Circuits}\ }\textbf {\bibinfo {volume} {34}},\ \bibinfo {pages} {586} (\bibinfo {year} {1999})}\BibitemShut {NoStop}%
\bibitem [{\citenamefont {Van~Delden}\ \emph {et~al.}(2019)\citenamefont {Van~Delden}, \citenamefont {Pohl}, \citenamefont {Aufinger}, \citenamefont {Baer},\ and\ \citenamefont {Musch}}]{van2019low}%
  \BibitemOpen
  \bibfield  {author} {\bibinfo {author} {\bibfnamefont {M.}~\bibnamefont {Van~Delden}}, \bibinfo {author} {\bibfnamefont {N.}~\bibnamefont {Pohl}}, \bibinfo {author} {\bibfnamefont {K.}~\bibnamefont {Aufinger}}, \bibinfo {author} {\bibfnamefont {C.}~\bibnamefont {Baer}},\ and\ \bibinfo {author} {\bibfnamefont {T.}~\bibnamefont {Musch}},\ }\bibfield  {title} {\bibinfo {title} {A low-noise transmission-type yttrium iron garnet tuned oscillator based on a sige mmic and bond-coupling operating up to 48 ghz},\ }\href {https://doi.org/https://doi.org/10.1109/TMTT.2019.2926293} {\bibfield  {journal} {\bibinfo  {journal} {IEEE Transactions on Microwave Theory and Techniques}\ }\textbf {\bibinfo {volume} {67}},\ \bibinfo {pages} {3973} (\bibinfo {year} {2019})}\BibitemShut {NoStop}%
\bibitem [{\citenamefont {Volkovskii}\ \emph {et~al.}(2005)\citenamefont {Volkovskii}, \citenamefont {Tsimring}, \citenamefont {Rulkov},\ and\ \citenamefont {Langmore}}]{volkovskii2005spread}%
  \BibitemOpen
  \bibfield  {author} {\bibinfo {author} {\bibfnamefont {A.}~\bibnamefont {Volkovskii}}, \bibinfo {author} {\bibfnamefont {L.~S.}\ \bibnamefont {Tsimring}}, \bibinfo {author} {\bibfnamefont {N.}~\bibnamefont {Rulkov}},\ and\ \bibinfo {author} {\bibfnamefont {I.}~\bibnamefont {Langmore}},\ }\bibfield  {title} {\bibinfo {title} {Spread spectrum communication system with chaotic frequency modulation},\ }\href@noop {} {\bibfield  {journal} {\bibinfo  {journal} {Chaos: An Interdisciplinary Journal of Nonlinear Science}\ }\textbf {\bibinfo {volume} {15}} (\bibinfo {year} {2005})}\BibitemShut {NoStop}%
\bibitem [{\citenamefont {Phan}\ \emph {et~al.}(2024)\citenamefont {Phan}, \citenamefont {Prasad}, \citenamefont {Hakam}, \citenamefont {Valli}, \citenamefont {Anghel}, \citenamefont {Benetti}, \citenamefont {Madhavan}, \citenamefont {Jenkins}, \citenamefont {Ferreira}, \citenamefont {Stiles} \emph {et~al.}}]{phan2024unbiased}%
  \BibitemOpen
  \bibfield  {author} {\bibinfo {author} {\bibfnamefont {N.-T.}\ \bibnamefont {Phan}}, \bibinfo {author} {\bibfnamefont {N.}~\bibnamefont {Prasad}}, \bibinfo {author} {\bibfnamefont {A.}~\bibnamefont {Hakam}}, \bibinfo {author} {\bibfnamefont {A.~S.~E.}\ \bibnamefont {Valli}}, \bibinfo {author} {\bibfnamefont {L.}~\bibnamefont {Anghel}}, \bibinfo {author} {\bibfnamefont {L.}~\bibnamefont {Benetti}}, \bibinfo {author} {\bibfnamefont {A.}~\bibnamefont {Madhavan}}, \bibinfo {author} {\bibfnamefont {A.~S.}\ \bibnamefont {Jenkins}}, \bibinfo {author} {\bibfnamefont {R.}~\bibnamefont {Ferreira}}, \bibinfo {author} {\bibfnamefont {M.~D.}\ \bibnamefont {Stiles}}, \emph {et~al.},\ }\bibfield  {title} {\bibinfo {title} {Unbiased random bitstream generation using injection-locked spin-torque nano-oscillators},\ }\href@noop {} {\bibfield  {journal} {\bibinfo  {journal} {Physical Review Applied}\ }\textbf {\bibinfo {volume} {21}},\ \bibinfo {pages} {034063} (\bibinfo {year} {2024})}\BibitemShut {NoStop}%
\bibitem [{\citenamefont {Dmitriev}\ \emph {et~al.}(2016)\citenamefont {Dmitriev}, \citenamefont {Efremova}, \citenamefont {Gerasimov},\ and\ \citenamefont {Itskov}}]{dmitriev2016radio}%
  \BibitemOpen
  \bibfield  {author} {\bibinfo {author} {\bibfnamefont {A.}~\bibnamefont {Dmitriev}}, \bibinfo {author} {\bibfnamefont {E.}~\bibnamefont {Efremova}}, \bibinfo {author} {\bibfnamefont {M.~Y.}\ \bibnamefont {Gerasimov}},\ and\ \bibinfo {author} {\bibfnamefont {V.}~\bibnamefont {Itskov}},\ }\bibfield  {title} {\bibinfo {title} {Radio lighting based on ultrawideband dynamic chaos generators},\ }\href@noop {} {\bibfield  {journal} {\bibinfo  {journal} {Journal of Communications Technology and Electronics}\ }\textbf {\bibinfo {volume} {61}},\ \bibinfo {pages} {1259} (\bibinfo {year} {2016})}\BibitemShut {NoStop}%
\bibitem [{\citenamefont {Dmitriev}\ and\ \citenamefont {Efremova}(2017)}]{dmitriev2017radio}%
  \BibitemOpen
  \bibfield  {author} {\bibinfo {author} {\bibfnamefont {A.}~\bibnamefont {Dmitriev}}\ and\ \bibinfo {author} {\bibfnamefont {E.}~\bibnamefont {Efremova}},\ }\bibfield  {title} {\bibinfo {title} {Radio-frequency illumination sources based on ultrawideband microgenerators of chaotic oscillations},\ }\href@noop {} {\bibfield  {journal} {\bibinfo  {journal} {Technical Physics Letters}\ }\textbf {\bibinfo {volume} {43}},\ \bibinfo {pages} {42} (\bibinfo {year} {2017})}\BibitemShut {NoStop}%
\bibitem [{\citenamefont {Liu}\ \emph {et~al.}(2020)\citenamefont {Liu}, \citenamefont {Cai}, \citenamefont {Zhang}, \citenamefont {Tovstopyat}, \citenamefont {Liu},\ and\ \citenamefont {Sun}}]{liu2020materials}%
  \BibitemOpen
  \bibfield  {author} {\bibinfo {author} {\bibfnamefont {Y.}~\bibnamefont {Liu}}, \bibinfo {author} {\bibfnamefont {Y.}~\bibnamefont {Cai}}, \bibinfo {author} {\bibfnamefont {Y.}~\bibnamefont {Zhang}}, \bibinfo {author} {\bibfnamefont {A.}~\bibnamefont {Tovstopyat}}, \bibinfo {author} {\bibfnamefont {S.}~\bibnamefont {Liu}},\ and\ \bibinfo {author} {\bibfnamefont {C.}~\bibnamefont {Sun}},\ }\bibfield  {title} {\bibinfo {title} {Materials, design, and characteristics of bulk acoustic wave resonator: A review},\ }\href {https://doi.org/https://doi.org/10.3390/mi11070630} {\bibfield  {journal} {\bibinfo  {journal} {Micromachines}\ }\textbf {\bibinfo {volume} {11}},\ \bibinfo {pages} {630} (\bibinfo {year} {2020})}\BibitemShut {NoStop}%
\bibitem [{\citenamefont {Zhao}\ \emph {et~al.}(2022)\citenamefont {Zhao}, \citenamefont {Asadi}, \citenamefont {Li}, \citenamefont {Chaudhuri}, \citenamefont {Nomoto}, \citenamefont {Xing}, \citenamefont {Hwang},\ and\ \citenamefont {Jena}}]{zhao2022x}%
  \BibitemOpen
  \bibfield  {author} {\bibinfo {author} {\bibfnamefont {W.}~\bibnamefont {Zhao}}, \bibinfo {author} {\bibfnamefont {M.~J.}\ \bibnamefont {Asadi}}, \bibinfo {author} {\bibfnamefont {L.}~\bibnamefont {Li}}, \bibinfo {author} {\bibfnamefont {R.}~\bibnamefont {Chaudhuri}}, \bibinfo {author} {\bibfnamefont {K.}~\bibnamefont {Nomoto}}, \bibinfo {author} {\bibfnamefont {H.~G.}\ \bibnamefont {Xing}}, \bibinfo {author} {\bibfnamefont {J.}~\bibnamefont {Hwang}},\ and\ \bibinfo {author} {\bibfnamefont {D.}~\bibnamefont {Jena}},\ }\bibfield  {title} {\bibinfo {title} {X-band epi-baw resonators},\ }\bibfield  {journal} {\bibinfo  {journal} {Journal of Applied Physics}\ }\textbf {\bibinfo {volume} {132}},\ \href {https://doi.org/https://doi.org/10.1063/5.0097458} {https://doi.org/10.1063/5.0097458} (\bibinfo {year} {2022})\BibitemShut {NoStop}%
\bibitem [{\citenamefont {Valle}\ and\ \citenamefont {Balram}(2021)}]{Valle2021HBARs}%
  \BibitemOpen
  \bibfield  {author} {\bibinfo {author} {\bibfnamefont {S.}~\bibnamefont {Valle}}\ and\ \bibinfo {author} {\bibfnamefont {K.~C.}\ \bibnamefont {Balram}},\ }\bibfield  {title} {\bibinfo {title} {High-overtone bulk acoustic resonators (hbar) as cryogenic high-frequency acousto-optic modulators},\ }in\ \href {https://doi.org/https://doi.org/10.1109/CLEO/Europe-EQEC52157.2021.9542604} {\emph {\bibinfo {booktitle} {2021 Conference on Lasers and Electro-Optics Europe \& European Quantum Electronics Conference (CLEO/Europe-EQEC)}}}\ (\bibinfo {year} {2021})\ pp.\ \bibinfo {pages} {1--1}\BibitemShut {NoStop}%
\bibitem [{\citenamefont {Aubin}\ \emph {et~al.}(2003)\citenamefont {Aubin}, \citenamefont {Zalalutdinov}, \citenamefont {Reichenbach}, \citenamefont {Houston}, \citenamefont {Zehnder}, \citenamefont {Parpia},\ and\ \citenamefont {Craighead}}]{aubin2003laser}%
  \BibitemOpen
  \bibfield  {author} {\bibinfo {author} {\bibfnamefont {K.~L.}\ \bibnamefont {Aubin}}, \bibinfo {author} {\bibfnamefont {M.}~\bibnamefont {Zalalutdinov}}, \bibinfo {author} {\bibfnamefont {R.~B.}\ \bibnamefont {Reichenbach}}, \bibinfo {author} {\bibfnamefont {B.~H.}\ \bibnamefont {Houston}}, \bibinfo {author} {\bibfnamefont {A.~T.}\ \bibnamefont {Zehnder}}, \bibinfo {author} {\bibfnamefont {J.~M.}\ \bibnamefont {Parpia}},\ and\ \bibinfo {author} {\bibfnamefont {H.~G.}\ \bibnamefont {Craighead}},\ }\bibfield  {title} {\bibinfo {title} {Laser annealing for high-q mems resonators},\ }in\ \href {https://doi.org/https://doi.org/10.1117/12.499109} {\emph {\bibinfo {booktitle} {Smart Sensors, Actuators, and MEMS}}},\ Vol.\ \bibinfo {volume} {5116}\ (\bibinfo {organization} {SPIE},\ \bibinfo {year} {2003})\ pp.\ \bibinfo {pages} {531--535}\BibitemShut {NoStop}%
\bibitem [{\citenamefont {Stassi}\ \emph {et~al.}(2021)\citenamefont {Stassi}, \citenamefont {Cooperstein}, \citenamefont {Tortello}, \citenamefont {Pirri}, \citenamefont {Magdassi},\ and\ \citenamefont {Ricciardi}}]{stassi2021reaching}%
  \BibitemOpen
  \bibfield  {author} {\bibinfo {author} {\bibfnamefont {S.}~\bibnamefont {Stassi}}, \bibinfo {author} {\bibfnamefont {I.}~\bibnamefont {Cooperstein}}, \bibinfo {author} {\bibfnamefont {M.}~\bibnamefont {Tortello}}, \bibinfo {author} {\bibfnamefont {C.~F.}\ \bibnamefont {Pirri}}, \bibinfo {author} {\bibfnamefont {S.}~\bibnamefont {Magdassi}},\ and\ \bibinfo {author} {\bibfnamefont {C.}~\bibnamefont {Ricciardi}},\ }\bibfield  {title} {\bibinfo {title} {Reaching silicon-based nems performances with 3d printed nanomechanical resonators},\ }\href {https://doi.org/https://doi.org/10.1038/s41467-021-26353-1} {\bibfield  {journal} {\bibinfo  {journal} {Nature Communications}\ }\textbf {\bibinfo {volume} {12}},\ \bibinfo {pages} {6080} (\bibinfo {year} {2021})}\BibitemShut {NoStop}%
\bibitem [{\citenamefont {Litvinenko}\ \emph {et~al.}(2021{\natexlab{a}})\citenamefont {Litvinenko}, \citenamefont {Khymyn}, \citenamefont {Tyberkevych}, \citenamefont {Tikhonov}, \citenamefont {Slavin},\ and\ \citenamefont {Nikitov}}]{litvinenko2021tunable}%
  \BibitemOpen
  \bibfield  {author} {\bibinfo {author} {\bibfnamefont {A.}~\bibnamefont {Litvinenko}}, \bibinfo {author} {\bibfnamefont {R.}~\bibnamefont {Khymyn}}, \bibinfo {author} {\bibfnamefont {V.}~\bibnamefont {Tyberkevych}}, \bibinfo {author} {\bibfnamefont {V.}~\bibnamefont {Tikhonov}}, \bibinfo {author} {\bibfnamefont {A.}~\bibnamefont {Slavin}},\ and\ \bibinfo {author} {\bibfnamefont {S.}~\bibnamefont {Nikitov}},\ }\bibfield  {title} {\bibinfo {title} {Tunable magnetoacoustic oscillator with low phase noise},\ }\href {https://doi.org/https://doi.org/10.1103/PhysRevApplied.15.034057} {\bibfield  {journal} {\bibinfo  {journal} {Physical Review Applied}\ }\textbf {\bibinfo {volume} {15}},\ \bibinfo {pages} {034057} (\bibinfo {year} {2021}{\natexlab{a}})}\BibitemShut {NoStop}%
\bibitem [{\citenamefont {Bankowski}\ \emph {et~al.}(2015)\citenamefont {Bankowski}, \citenamefont {Meitzler}, \citenamefont {Khymyn}, \citenamefont {Tiberkevich}, \citenamefont {Slavin},\ and\ \citenamefont {Tang}}]{Bankowski2015magnon}%
  \BibitemOpen
  \bibfield  {author} {\bibinfo {author} {\bibfnamefont {E.}~\bibnamefont {Bankowski}}, \bibinfo {author} {\bibfnamefont {T.}~\bibnamefont {Meitzler}}, \bibinfo {author} {\bibfnamefont {R.~S.}\ \bibnamefont {Khymyn}}, \bibinfo {author} {\bibfnamefont {V.~S.}\ \bibnamefont {Tiberkevich}}, \bibinfo {author} {\bibfnamefont {A.~N.}\ \bibnamefont {Slavin}},\ and\ \bibinfo {author} {\bibfnamefont {H.~X.}\ \bibnamefont {Tang}},\ }\bibfield  {title} {\bibinfo {title} {Magnonic crystal as a delay line for low-noise auto-oscillators},\ }\href {https://doi.org/https://doi.org/10.1063/1.4931758} {\bibfield  {journal} {\bibinfo  {journal} {Applied Physics Letters}\ }\textbf {\bibinfo {volume} {107}},\ \bibinfo {pages} {122409} (\bibinfo {year} {2015})}\BibitemShut {NoStop}%
\bibitem [{\citenamefont {Dai}\ \emph {et~al.}(2020)\citenamefont {Dai}, \citenamefont {Bhave},\ and\ \citenamefont {Wang}}]{dai2020octave}%
  \BibitemOpen
  \bibfield  {author} {\bibinfo {author} {\bibfnamefont {S.}~\bibnamefont {Dai}}, \bibinfo {author} {\bibfnamefont {S.~A.}\ \bibnamefont {Bhave}},\ and\ \bibinfo {author} {\bibfnamefont {R.}~\bibnamefont {Wang}},\ }\bibfield  {title} {\bibinfo {title} {Octave-tunable magnetostatic wave yig resonators on a chip},\ }\href {https://doi.org/https://doi.org/10.1109/TUFFC.2020.3000055} {\bibfield  {journal} {\bibinfo  {journal} {IEEE Transactions on Ultrasonics, Ferroelectrics, and Frequency Control}\ }\textbf {\bibinfo {volume} {67}},\ \bibinfo {pages} {2454} (\bibinfo {year} {2020})}\BibitemShut {NoStop}%
\bibitem [{\citenamefont {Gevorkyan}\ \emph {et~al.}(2021)\citenamefont {Gevorkyan}, \citenamefont {Kochemasov}, \citenamefont {Safin},\ and\ \citenamefont {Chenakin}}]{gevorkyan2021ferrite}%
  \BibitemOpen
  \bibfield  {author} {\bibinfo {author} {\bibfnamefont {V.}~\bibnamefont {Gevorkyan}}, \bibinfo {author} {\bibfnamefont {V.}~\bibnamefont {Kochemasov}}, \bibinfo {author} {\bibfnamefont {A.}~\bibnamefont {Safin}},\ and\ \bibinfo {author} {\bibfnamefont {A.}~\bibnamefont {Chenakin}},\ }\bibfield  {title} {\bibinfo {title} {Ferrite-based microwave oscillators},\ }in\ \href@noop {} {\emph {\bibinfo {booktitle} {2021 Systems of Signal Synchronization, Generating and Processing in Telecommunications (SYNCHROINFO}}}\ (\bibinfo {organization} {IEEE},\ \bibinfo {year} {2021})\ pp.\ \bibinfo {pages} {1--4}\BibitemShut {NoStop}%
\bibitem [{\citenamefont {Nikitin}\ \emph {et~al.}(2024)\citenamefont {Nikitin}, \citenamefont {Tatsenko}, \citenamefont {Kostylev},\ and\ \citenamefont {Ustinov}}]{nikitin2024microwave}%
  \BibitemOpen
  \bibfield  {author} {\bibinfo {author} {\bibfnamefont {A.~A.}\ \bibnamefont {Nikitin}}, \bibinfo {author} {\bibfnamefont {I.~Y.}\ \bibnamefont {Tatsenko}}, \bibinfo {author} {\bibfnamefont {M.~P.}\ \bibnamefont {Kostylev}},\ and\ \bibinfo {author} {\bibfnamefont {A.~B.}\ \bibnamefont {Ustinov}},\ }\bibfield  {title} {\bibinfo {title} {Microwave magnonic micro-oscillator based on a nm-thick yig film},\ }\bibfield  {journal} {\bibinfo  {journal} {Journal of Applied Physics}\ }\textbf {\bibinfo {volume} {135}},\ \href {https://doi.org/https://doi.org/10.1063/5.0200249} {https://doi.org/10.1063/5.0200249} (\bibinfo {year} {2024})\BibitemShut {NoStop}%
\bibitem [{\citenamefont {Serga}\ \emph {et~al.}(2010)\citenamefont {Serga}, \citenamefont {Chumak},\ and\ \citenamefont {Hillebrands}}]{serga2010yig}%
  \BibitemOpen
  \bibfield  {author} {\bibinfo {author} {\bibfnamefont {A.}~\bibnamefont {Serga}}, \bibinfo {author} {\bibfnamefont {A.}~\bibnamefont {Chumak}},\ and\ \bibinfo {author} {\bibfnamefont {B.}~\bibnamefont {Hillebrands}},\ }\bibfield  {title} {\bibinfo {title} {Yig magnonics},\ }\href {https://doi.org/https://doi.org/10.1088/0022-3727/43/26/264002} {\bibfield  {journal} {\bibinfo  {journal} {Journal of Physics D: Applied Physics}\ }\textbf {\bibinfo {volume} {43}},\ \bibinfo {pages} {264002} (\bibinfo {year} {2010})}\BibitemShut {NoStop}%
\bibitem [{\citenamefont {Lakin}\ \emph {et~al.}(1993)\citenamefont {Lakin}, \citenamefont {Kline},\ and\ \citenamefont {McCarron}}]{lakin1993high}%
  \BibitemOpen
  \bibfield  {author} {\bibinfo {author} {\bibfnamefont {K.~M.}\ \bibnamefont {Lakin}}, \bibinfo {author} {\bibfnamefont {G.~R.}\ \bibnamefont {Kline}},\ and\ \bibinfo {author} {\bibfnamefont {K.~T.}\ \bibnamefont {McCarron}},\ }\bibfield  {title} {\bibinfo {title} {High-q microwave acoustic resonators and filters},\ }\href {https://doi.org/https://doi.org/10.1109/22.260698} {\bibfield  {journal} {\bibinfo  {journal} {IEEE transactions on microwave theory and techniques}\ }\textbf {\bibinfo {volume} {41}},\ \bibinfo {pages} {2139} (\bibinfo {year} {1993})}\BibitemShut {NoStop}%
\bibitem [{\citenamefont {Litvinenko}\ \emph {et~al.}(2015)\citenamefont {Litvinenko}, \citenamefont {Sadovnikov}, \citenamefont {Tikhonov},\ and\ \citenamefont {Nikitov}}]{litvinenko2015brillouin}%
  \BibitemOpen
  \bibfield  {author} {\bibinfo {author} {\bibfnamefont {A.~N.}\ \bibnamefont {Litvinenko}}, \bibinfo {author} {\bibfnamefont {A.~V.}\ \bibnamefont {Sadovnikov}}, \bibinfo {author} {\bibfnamefont {V.~V.}\ \bibnamefont {Tikhonov}},\ and\ \bibinfo {author} {\bibfnamefont {S.~A.}\ \bibnamefont {Nikitov}},\ }\bibfield  {title} {\bibinfo {title} {Brillouin light scattering spectroscopy of magneto-acoustic resonances in a thin-film garnet resonator},\ }\href {https://doi.org/https://doi.org/10.1109/LMAG.2015.2494008} {\bibfield  {journal} {\bibinfo  {journal} {IEEE Magnetics Letters}\ }\textbf {\bibinfo {volume} {6}},\ \bibinfo {pages} {1} (\bibinfo {year} {2015})}\BibitemShut {NoStop}%
\bibitem [{\citenamefont {Tikhonov}\ \emph {et~al.}(2016)\citenamefont {Tikhonov}, \citenamefont {Litvinenko}, \citenamefont {Sadovnikov},\ and\ \citenamefont {Nikitov}}]{tikhonov2016brillouin}%
  \BibitemOpen
  \bibfield  {author} {\bibinfo {author} {\bibfnamefont {V.}~\bibnamefont {Tikhonov}}, \bibinfo {author} {\bibfnamefont {A.}~\bibnamefont {Litvinenko}}, \bibinfo {author} {\bibfnamefont {A.}~\bibnamefont {Sadovnikov}},\ and\ \bibinfo {author} {\bibfnamefont {S.}~\bibnamefont {Nikitov}},\ }\bibfield  {title} {\bibinfo {title} {Brillouin spectroscopy of nonlinear magnetoacoustic resonances in a layered yig/ggg structure},\ }\href {https://doi.org/https://doi.org/10.3103/S1062873816100208} {\bibfield  {journal} {\bibinfo  {journal} {Bulletin of the Russian Academy of Sciences: Physics}\ }\textbf {\bibinfo {volume} {80}},\ \bibinfo {pages} {1242} (\bibinfo {year} {2016})}\BibitemShut {NoStop}%
\bibitem [{\citenamefont {Comstock}(1965)}]{comstock1965magnetoelastic}%
  \BibitemOpen
  \bibfield  {author} {\bibinfo {author} {\bibfnamefont {R.}~\bibnamefont {Comstock}},\ }\bibfield  {title} {\bibinfo {title} {Magnetoelastic coupling constants of the ferrites and garnets},\ }\href {https://doi.org/https://doi.org/10.1016/B978-0-12-395669-9.50011-4} {\bibfield  {journal} {\bibinfo  {journal} {Proceedings of the IEEE}\ }\textbf {\bibinfo {volume} {53}},\ \bibinfo {pages} {1508} (\bibinfo {year} {1965})}\BibitemShut {NoStop}%
\bibitem [{\citenamefont {Gulyaev}\ \emph {et~al.}(1981)\citenamefont {Gulyaev}, \citenamefont {Zilberman}, \citenamefont {Kazakov}, \citenamefont {Nam}, \citenamefont {Tikhonov}, \citenamefont {Filimonov},\ and\ \citenamefont {Khe}}]{gulyaev1981observation}%
  \BibitemOpen
  \bibfield  {author} {\bibinfo {author} {\bibfnamefont {Y.~V.}\ \bibnamefont {Gulyaev}}, \bibinfo {author} {\bibfnamefont {P.}~\bibnamefont {Zilberman}}, \bibinfo {author} {\bibfnamefont {G.}~\bibnamefont {Kazakov}}, \bibinfo {author} {\bibfnamefont {B.}~\bibnamefont {Nam}}, \bibinfo {author} {\bibfnamefont {V.}~\bibnamefont {Tikhonov}}, \bibinfo {author} {\bibfnamefont {Y.}~\bibnamefont {Filimonov}},\ and\ \bibinfo {author} {\bibfnamefont {A.}~\bibnamefont {Khe}},\ }\bibfield  {title} {\bibinfo {title} {Observation of fast magnetoelastic waves in thin yttrium-iron garnet wafers and epitaxial films},\ }\href {https://doi.org/https://ui.adsabs.harvard.edu/abs/1981JETPL..34..477G} {\bibfield  {journal} {\bibinfo  {journal} {JETP Let.}\ }\textbf {\bibinfo {volume} {34}},\ \bibinfo {pages} {477} (\bibinfo {year} {1981})}\BibitemShut {NoStop}%
\bibitem [{\citenamefont {Kazakov}\ \emph {et~al.}(1983)\citenamefont {Kazakov}, \citenamefont {Tikhonov},\ and\ \citenamefont {Zil'berman}}]{kazakov1983magneto}%
  \BibitemOpen
  \bibfield  {author} {\bibinfo {author} {\bibfnamefont {G.}~\bibnamefont {Kazakov}}, \bibinfo {author} {\bibfnamefont {V.}~\bibnamefont {Tikhonov}},\ and\ \bibinfo {author} {\bibfnamefont {P.~E.}\ \bibnamefont {Zil'berman}},\ }\bibfield  {title} {\bibinfo {title} {Magneto-dipole and elastic wave resonance interaction in yig plates and films},\ }\href {https://doi.org/https://doi.org/10.1134/1.1129801} {\bibfield  {journal} {\bibinfo  {journal} {Fizika Tverdogo Tela}\ }\textbf {\bibinfo {volume} {25}},\ \bibinfo {pages} {2307} (\bibinfo {year} {1983})}\BibitemShut {NoStop}%
\bibitem [{\citenamefont {Zilberman}\ \emph {et~al.}(1985)\citenamefont {Zilberman}, \citenamefont {Kazakov},\ and\ \citenamefont {Tikhonov}}]{zilberman1985self}%
  \BibitemOpen
  \bibfield  {author} {\bibinfo {author} {\bibfnamefont {P.~E.}\ \bibnamefont {Zilberman}}, \bibinfo {author} {\bibfnamefont {G.}~\bibnamefont {Kazakov}},\ and\ \bibinfo {author} {\bibfnamefont {V.~V.}\ \bibnamefont {Tikhonov}},\ }\bibfield  {title} {\bibinfo {title} {Self-modulation of fast magnetoelastic waves in yttrium iron garnet films},\ }\href {https://doi.org/https://www.mathnet.ru/eng/pjtf/v11/i13/p769} {\bibfield  {journal} {\bibinfo  {journal} {Pisma v Zhurnal Tekhnischeskoi Fiziki}\ }\textbf {\bibinfo {volume} {11}},\ \bibinfo {pages} {769} (\bibinfo {year} {1985})}\BibitemShut {NoStop}%
\bibitem [{\citenamefont {Gulyaev}\ and\ \citenamefont {Zil'Berman}(1988)}]{gulyaev1988magnetoelastic}%
  \BibitemOpen
  \bibfield  {author} {\bibinfo {author} {\bibfnamefont {Y.~V.}\ \bibnamefont {Gulyaev}}\ and\ \bibinfo {author} {\bibfnamefont {P.}~\bibnamefont {Zil'Berman}},\ }\bibfield  {title} {\bibinfo {title} {Magnetoelastic waves in ferromagnet plates and films},\ }\href {https://doi.org/https://doi.org/10.1007/BF00893540} {\bibfield  {journal} {\bibinfo  {journal} {Soviet Physics Journal}\ }\textbf {\bibinfo {volume} {31}},\ \bibinfo {pages} {860} (\bibinfo {year} {1988})}\BibitemShut {NoStop}%
\bibitem [{\citenamefont {Verba}\ \emph {et~al.}(2018)\citenamefont {Verba}, \citenamefont {Lisenkov}, \citenamefont {Krivorotov}, \citenamefont {Tiberkevich},\ and\ \citenamefont {Slavin}}]{verba2018nonreciprocal}%
  \BibitemOpen
  \bibfield  {author} {\bibinfo {author} {\bibfnamefont {R.}~\bibnamefont {Verba}}, \bibinfo {author} {\bibfnamefont {I.}~\bibnamefont {Lisenkov}}, \bibinfo {author} {\bibfnamefont {I.}~\bibnamefont {Krivorotov}}, \bibinfo {author} {\bibfnamefont {V.}~\bibnamefont {Tiberkevich}},\ and\ \bibinfo {author} {\bibfnamefont {A.}~\bibnamefont {Slavin}},\ }\bibfield  {title} {\bibinfo {title} {Nonreciprocal surface acoustic waves in multilayers with magnetoelastic and interfacial dzyaloshinskii-moriya interactions},\ }\href {https://doi.org/10.1103/PhysRevApplied.9.064014} {\bibfield  {journal} {\bibinfo  {journal} {Phys. Rev. Appl.}\ }\textbf {\bibinfo {volume} {9}},\ \bibinfo {pages} {064014} (\bibinfo {year} {2018})}\BibitemShut {NoStop}%
\bibitem [{\citenamefont {Lisenkov}\ \emph {et~al.}(2019)\citenamefont {Lisenkov}, \citenamefont {Jander},\ and\ \citenamefont {Dhagat}}]{lisenkov2019magnetoelastic}%
  \BibitemOpen
  \bibfield  {author} {\bibinfo {author} {\bibfnamefont {I.}~\bibnamefont {Lisenkov}}, \bibinfo {author} {\bibfnamefont {A.}~\bibnamefont {Jander}},\ and\ \bibinfo {author} {\bibfnamefont {P.}~\bibnamefont {Dhagat}},\ }\bibfield  {title} {\bibinfo {title} {Magnetoelastic parametric instabilities of localized spin waves induced by traveling elastic waves},\ }\href {https://doi.org/10.1103/PhysRevB.99.184433} {\bibfield  {journal} {\bibinfo  {journal} {Phys. Rev. B}\ }\textbf {\bibinfo {volume} {99}},\ \bibinfo {pages} {184433} (\bibinfo {year} {2019})}\BibitemShut {NoStop}%
\bibitem [{\citenamefont {Brataas}\ \emph {et~al.}(2020)\citenamefont {Brataas}, \citenamefont {van Wees}, \citenamefont {Klein}, \citenamefont {de~Loubens},\ and\ \citenamefont {Viret}}]{brataas2020spin}%
  \BibitemOpen
  \bibfield  {author} {\bibinfo {author} {\bibfnamefont {A.}~\bibnamefont {Brataas}}, \bibinfo {author} {\bibfnamefont {B.}~\bibnamefont {van Wees}}, \bibinfo {author} {\bibfnamefont {O.}~\bibnamefont {Klein}}, \bibinfo {author} {\bibfnamefont {G.}~\bibnamefont {de~Loubens}},\ and\ \bibinfo {author} {\bibfnamefont {M.}~\bibnamefont {Viret}},\ }\bibfield  {title} {\bibinfo {title} {Spin insulatronics},\ }\href {https://doi.org/10.1016/j.physrep.2020.08.006} {\bibfield  {journal} {\bibinfo  {journal} {Physics Reports}\ }\textbf {\bibinfo {volume} {885}},\ \bibinfo {pages} {1} (\bibinfo {year} {2020})}\BibitemShut {NoStop}%
\bibitem [{\citenamefont {An}\ \emph {et~al.}(2020)\citenamefont {An}, \citenamefont {Litvinenko}, \citenamefont {Kohno}, \citenamefont {Fuad}, \citenamefont {Naletov}, \citenamefont {Vila}, \citenamefont {Ebels}, \citenamefont {de~Loubens}, \citenamefont {Hurdequint}, \citenamefont {Beaulieu}, \citenamefont {Ben~Youssef}, \citenamefont {Vukadinovic}, \citenamefont {Bauer}, \citenamefont {Slavin}, \citenamefont {Tiberkevich},\ and\ \citenamefont {Klein}}]{An2020}%
  \BibitemOpen
  \bibfield  {author} {\bibinfo {author} {\bibfnamefont {K.}~\bibnamefont {An}}, \bibinfo {author} {\bibfnamefont {A.~N.}\ \bibnamefont {Litvinenko}}, \bibinfo {author} {\bibfnamefont {R.}~\bibnamefont {Kohno}}, \bibinfo {author} {\bibfnamefont {A.~A.}\ \bibnamefont {Fuad}}, \bibinfo {author} {\bibfnamefont {V.~V.}\ \bibnamefont {Naletov}}, \bibinfo {author} {\bibfnamefont {L.}~\bibnamefont {Vila}}, \bibinfo {author} {\bibfnamefont {U.}~\bibnamefont {Ebels}}, \bibinfo {author} {\bibfnamefont {G.}~\bibnamefont {de~Loubens}}, \bibinfo {author} {\bibfnamefont {H.}~\bibnamefont {Hurdequint}}, \bibinfo {author} {\bibfnamefont {N.}~\bibnamefont {Beaulieu}}, \bibinfo {author} {\bibfnamefont {J.}~\bibnamefont {Ben~Youssef}}, \bibinfo {author} {\bibfnamefont {N.}~\bibnamefont {Vukadinovic}}, \bibinfo {author} {\bibfnamefont {G.~E.~W.}\ \bibnamefont {Bauer}}, \bibinfo {author} {\bibfnamefont {A.~N.}\ \bibnamefont {Slavin}}, \bibinfo {author} {\bibfnamefont {V.~S.}\ \bibnamefont {Tiberkevich}},\ and\ \bibinfo {author}
  {\bibfnamefont {O.}~\bibnamefont {Klein}},\ }\bibfield  {title} {\bibinfo {title} {Coherent long-range transfer of angular momentum between magnon kittel modes by phonons},\ }\href {https://doi.org/10.1103/PhysRevB.101.060407} {\bibfield  {journal} {\bibinfo  {journal} {Phys. Rev. B}\ }\textbf {\bibinfo {volume} {101}},\ \bibinfo {pages} {060407(R)} (\bibinfo {year} {2020})}\BibitemShut {NoStop}%
\bibitem [{\citenamefont {An}\ \emph {et~al.}(2022)\citenamefont {An}, \citenamefont {Kohno}, \citenamefont {Litvinenko}, \citenamefont {Seeger}, \citenamefont {Naletov}, \citenamefont {Vila}, \citenamefont {de~Loubens}, \citenamefont {Ben~Youssef}, \citenamefont {Vukadinovic}, \citenamefont {Bauer}, \citenamefont {Slavin}, \citenamefont {Tiberkevich},\ and\ \citenamefont {Klein}}]{an2022bright}%
  \BibitemOpen
  \bibfield  {author} {\bibinfo {author} {\bibfnamefont {K.}~\bibnamefont {An}}, \bibinfo {author} {\bibfnamefont {R.}~\bibnamefont {Kohno}}, \bibinfo {author} {\bibfnamefont {A.~N.}\ \bibnamefont {Litvinenko}}, \bibinfo {author} {\bibfnamefont {R.~L.}\ \bibnamefont {Seeger}}, \bibinfo {author} {\bibfnamefont {V.~V.}\ \bibnamefont {Naletov}}, \bibinfo {author} {\bibfnamefont {L.}~\bibnamefont {Vila}}, \bibinfo {author} {\bibfnamefont {G.}~\bibnamefont {de~Loubens}}, \bibinfo {author} {\bibfnamefont {J.}~\bibnamefont {Ben~Youssef}}, \bibinfo {author} {\bibfnamefont {N.}~\bibnamefont {Vukadinovic}}, \bibinfo {author} {\bibfnamefont {G.~E.~W.}\ \bibnamefont {Bauer}}, \bibinfo {author} {\bibfnamefont {A.~N.}\ \bibnamefont {Slavin}}, \bibinfo {author} {\bibfnamefont {V.~S.}\ \bibnamefont {Tiberkevich}},\ and\ \bibinfo {author} {\bibfnamefont {O.}~\bibnamefont {Klein}},\ }\bibfield  {title} {\bibinfo {title} {Bright and dark states of two distant macrospins strongly coupled by phonons},\ }\href
  {https://doi.org/https://doi.org/10.1103/PhysRevX.12.011060} {\bibfield  {journal} {\bibinfo  {journal} {Phys. Rev. X}\ }\textbf {\bibinfo {volume} {12}},\ \bibinfo {pages} {011060} (\bibinfo {year} {2022})}\BibitemShut {NoStop}%
\bibitem [{\citenamefont {Litvinenko}\ \emph {et~al.}(2018)\citenamefont {Litvinenko}, \citenamefont {Grishin}, \citenamefont {Sharaevskii}, \citenamefont {Tikhonov},\ and\ \citenamefont {Nikitov}}]{litvinenko2018chaotic}%
  \BibitemOpen
  \bibfield  {author} {\bibinfo {author} {\bibfnamefont {A.}~\bibnamefont {Litvinenko}}, \bibinfo {author} {\bibfnamefont {S.}~\bibnamefont {Grishin}}, \bibinfo {author} {\bibfnamefont {Y.~P.}\ \bibnamefont {Sharaevskii}}, \bibinfo {author} {\bibfnamefont {V.}~\bibnamefont {Tikhonov}},\ and\ \bibinfo {author} {\bibfnamefont {S.}~\bibnamefont {Nikitov}},\ }\bibfield  {title} {\bibinfo {title} {A chaotic magnetoacoustic oscillator with delay and bistability},\ }\href {https://doi.org/https://doi.org/10.1134/S1063785018030215} {\bibfield  {journal} {\bibinfo  {journal} {Technical Physics Letters}\ }\textbf {\bibinfo {volume} {44}},\ \bibinfo {pages} {263} (\bibinfo {year} {2018})}\BibitemShut {NoStop}%
\bibitem [{\citenamefont {Ustinov}\ \emph {et~al.}(2009)\citenamefont {Ustinov}, \citenamefont {Kalinikos}, \citenamefont {Demidov},\ and\ \citenamefont {Demokritov}}]{ustinov2009generation}%
  \BibitemOpen
  \bibfield  {author} {\bibinfo {author} {\bibfnamefont {A.~B.}\ \bibnamefont {Ustinov}}, \bibinfo {author} {\bibfnamefont {B.~A.}\ \bibnamefont {Kalinikos}}, \bibinfo {author} {\bibfnamefont {V.~E.}\ \bibnamefont {Demidov}},\ and\ \bibinfo {author} {\bibfnamefont {S.~O.}\ \bibnamefont {Demokritov}},\ }\bibfield  {title} {\bibinfo {title} {Generation of dense spin-wave soliton trains in active ring resonators},\ }\href {https://doi.org/https://doi.org/10.1103/PhysRevB.80.052405} {\bibfield  {journal} {\bibinfo  {journal} {Physical Review B—Condensed Matter and Materials Physics}\ }\textbf {\bibinfo {volume} {80}},\ \bibinfo {pages} {052405} (\bibinfo {year} {2009})}\BibitemShut {NoStop}%
\bibitem [{\citenamefont {Ustinov}\ \emph {et~al.}(2014)\citenamefont {Ustinov}, \citenamefont {Kondrashov}, \citenamefont {Nikitin},\ and\ \citenamefont {Kalinikos}}]{ustinov2014self}%
  \BibitemOpen
  \bibfield  {author} {\bibinfo {author} {\bibfnamefont {A.~B.}\ \bibnamefont {Ustinov}}, \bibinfo {author} {\bibfnamefont {A.~V.}\ \bibnamefont {Kondrashov}}, \bibinfo {author} {\bibfnamefont {A.~A.}\ \bibnamefont {Nikitin}},\ and\ \bibinfo {author} {\bibfnamefont {B.~A.}\ \bibnamefont {Kalinikos}},\ }\bibfield  {title} {\bibinfo {title} {Self-generation and management of spin-electromagnetic wave solitons and chaos},\ }\bibfield  {journal} {\bibinfo  {journal} {Applied Physics Letters}\ }\textbf {\bibinfo {volume} {104}},\ \href {https://doi.org/https://doi.org/10.1063/1.4881889} {https://doi.org/10.1063/1.4881889} (\bibinfo {year} {2014})\BibitemShut {NoStop}%
\bibitem [{\citenamefont {Bir}\ \emph {et~al.}(2020)\citenamefont {Bir}, \citenamefont {Grishin}, \citenamefont {Moskalenko}, \citenamefont {Pavlov}, \citenamefont {Zhuravlev},\ and\ \citenamefont {Ruiz}}]{bir2020experimental}%
  \BibitemOpen
  \bibfield  {author} {\bibinfo {author} {\bibfnamefont {A.~S.}\ \bibnamefont {Bir}}, \bibinfo {author} {\bibfnamefont {S.~V.}\ \bibnamefont {Grishin}}, \bibinfo {author} {\bibfnamefont {O.~I.}\ \bibnamefont {Moskalenko}}, \bibinfo {author} {\bibfnamefont {A.~N.}\ \bibnamefont {Pavlov}}, \bibinfo {author} {\bibfnamefont {M.~O.}\ \bibnamefont {Zhuravlev}},\ and\ \bibinfo {author} {\bibfnamefont {D.~O.}\ \bibnamefont {Ruiz}},\ }\bibfield  {title} {\bibinfo {title} {Experimental observation of ultrashort hyperchaotic dark multisoliton complexes in a magnonic active ring resonator},\ }\href {https://doi.org/https://doi.org/10.1103/PhysRevLett.125.083903} {\bibfield  {journal} {\bibinfo  {journal} {Physical Review Letters}\ }\textbf {\bibinfo {volume} {125}},\ \bibinfo {pages} {083903} (\bibinfo {year} {2020})}\BibitemShut {NoStop}%
\bibitem [{\citenamefont {Bir}\ \emph {et~al.}(2024)\citenamefont {Bir}, \citenamefont {Grishin}, \citenamefont {Grachev}, \citenamefont {Moskalenko}, \citenamefont {Pavlov}, \citenamefont {Romanenko}, \citenamefont {Skorokhodov},\ and\ \citenamefont {Nikitov}}]{bir2024direct}%
  \BibitemOpen
  \bibfield  {author} {\bibinfo {author} {\bibfnamefont {A.~S.}\ \bibnamefont {Bir}}, \bibinfo {author} {\bibfnamefont {S.~V.}\ \bibnamefont {Grishin}}, \bibinfo {author} {\bibfnamefont {A.~A.}\ \bibnamefont {Grachev}}, \bibinfo {author} {\bibfnamefont {O.~I.}\ \bibnamefont {Moskalenko}}, \bibinfo {author} {\bibfnamefont {A.~N.}\ \bibnamefont {Pavlov}}, \bibinfo {author} {\bibfnamefont {D.~V.}\ \bibnamefont {Romanenko}}, \bibinfo {author} {\bibfnamefont {V.~N.}\ \bibnamefont {Skorokhodov}},\ and\ \bibinfo {author} {\bibfnamefont {S.~A.}\ \bibnamefont {Nikitov}},\ }\bibfield  {title} {\bibinfo {title} {Direct electric current control of hyperchaotic packets of dissipative dark envelope solitons in a magnonic crystal active ring resonator},\ }\href {https://doi.org/https://doi.org/10.1103/PhysRevApplied.21.044008} {\bibfield  {journal} {\bibinfo  {journal} {Physical Review Applied}\ }\textbf {\bibinfo {volume} {21}},\ \bibinfo {pages} {044008} (\bibinfo {year} {2024})}\BibitemShut {NoStop}%
\bibitem [{\citenamefont {Massouras}\ \emph {et~al.}(2024)\citenamefont {Massouras}, \citenamefont {Perna}, \citenamefont {d'Aquino}, \citenamefont {Serpico},\ and\ \citenamefont {Kim}}]{massouras2024mode}%
  \BibitemOpen
  \bibfield  {author} {\bibinfo {author} {\bibfnamefont {M.}~\bibnamefont {Massouras}}, \bibinfo {author} {\bibfnamefont {S.}~\bibnamefont {Perna}}, \bibinfo {author} {\bibfnamefont {M.}~\bibnamefont {d'Aquino}}, \bibinfo {author} {\bibfnamefont {C.}~\bibnamefont {Serpico}},\ and\ \bibinfo {author} {\bibfnamefont {J.-V.}\ \bibnamefont {Kim}},\ }\bibfield  {title} {\bibinfo {title} {Mode-resolved micromagnetics study of parametric spin wave excitation in thin-film disks},\ }\href {https://doi.org/https://doi.org/10.1103/PhysRevB.110.064435} {\bibfield  {journal} {\bibinfo  {journal} {Physical Review B}\ }\textbf {\bibinfo {volume} {110}},\ \bibinfo {pages} {064435} (\bibinfo {year} {2024})}\BibitemShut {NoStop}%
\bibitem [{\citenamefont {Srivastava}\ \emph {et~al.}(2023)\citenamefont {Srivastava}, \citenamefont {Merbouche}, \citenamefont {Ngouagnia~Yemeli}, \citenamefont {Beaulieu}, \citenamefont {Ben~Youssef}, \citenamefont {Mu\~noz}, \citenamefont {Che}, \citenamefont {Bortolotti}, \citenamefont {Cros}, \citenamefont {Klein}, \citenamefont {Sangiao}, \citenamefont {De~Teresa}, \citenamefont {Demokritov}, \citenamefont {Demidov}, \citenamefont {Anane}, \citenamefont {Serpico}, \citenamefont {d'Aquino},\ and\ \citenamefont {de~Loubens}}]{srivastava2023YIGeigen}%
  \BibitemOpen
  \bibfield  {author} {\bibinfo {author} {\bibfnamefont {T.}~\bibnamefont {Srivastava}}, \bibinfo {author} {\bibfnamefont {H.}~\bibnamefont {Merbouche}}, \bibinfo {author} {\bibfnamefont {I.}~\bibnamefont {Ngouagnia~Yemeli}}, \bibinfo {author} {\bibfnamefont {N.}~\bibnamefont {Beaulieu}}, \bibinfo {author} {\bibfnamefont {J.}~\bibnamefont {Ben~Youssef}}, \bibinfo {author} {\bibfnamefont {M.}~\bibnamefont {Mu\~noz}}, \bibinfo {author} {\bibfnamefont {P.}~\bibnamefont {Che}}, \bibinfo {author} {\bibfnamefont {P.}~\bibnamefont {Bortolotti}}, \bibinfo {author} {\bibfnamefont {V.}~\bibnamefont {Cros}}, \bibinfo {author} {\bibfnamefont {O.}~\bibnamefont {Klein}}, \bibinfo {author} {\bibfnamefont {S.}~\bibnamefont {Sangiao}}, \bibinfo {author} {\bibfnamefont {J.}~\bibnamefont {De~Teresa}}, \bibinfo {author} {\bibfnamefont {S.}~\bibnamefont {Demokritov}}, \bibinfo {author} {\bibfnamefont {V.}~\bibnamefont {Demidov}}, \bibinfo {author} {\bibfnamefont {A.}~\bibnamefont {Anane}}, \bibinfo {author} {\bibfnamefont
  {C.}~\bibnamefont {Serpico}}, \bibinfo {author} {\bibfnamefont {M.}~\bibnamefont {d'Aquino}},\ and\ \bibinfo {author} {\bibfnamefont {G.}~\bibnamefont {de~Loubens}},\ }\bibfield  {title} {\bibinfo {title} {Identification of a large number of spin-wave eigenmodes excited by parametric pumping in yttrium iron garnet microdisks},\ }\href {https://doi.org/10.1103/PhysRevApplied.19.064078} {\bibfield  {journal} {\bibinfo  {journal} {Phys. Rev. Appl.}\ }\textbf {\bibinfo {volume} {19}},\ \bibinfo {pages} {064078} (\bibinfo {year} {2023})}\BibitemShut {NoStop}%
\bibitem [{\citenamefont {Demokritov}\ \emph {et~al.}(2001)\citenamefont {Demokritov}, \citenamefont {Hillebrands},\ and\ \citenamefont {Slavin}}]{demokritov2001brillouin}%
  \BibitemOpen
  \bibfield  {author} {\bibinfo {author} {\bibfnamefont {S.~O.}\ \bibnamefont {Demokritov}}, \bibinfo {author} {\bibfnamefont {B.}~\bibnamefont {Hillebrands}},\ and\ \bibinfo {author} {\bibfnamefont {A.~N.}\ \bibnamefont {Slavin}},\ }\bibfield  {title} {\bibinfo {title} {Brillouin light scattering studies of confined spin waves: linear and nonlinear confinement},\ }\href {https://doi.org/https://doi.org/10.1016/S0370-1573(00)00116-2} {\bibfield  {journal} {\bibinfo  {journal} {Physics Reports}\ }\textbf {\bibinfo {volume} {348}},\ \bibinfo {pages} {441} (\bibinfo {year} {2001})}\BibitemShut {NoStop}%
\bibitem [{\citenamefont {An}\ \emph {et~al.}(2024)\citenamefont {An}, \citenamefont {Xu}, \citenamefont {Mucchietto}, \citenamefont {Kim}, \citenamefont {Moon}, \citenamefont {Hwang},\ and\ \citenamefont {Grundler}}]{an2024emergent}%
  \BibitemOpen
  \bibfield  {author} {\bibinfo {author} {\bibfnamefont {K.}~\bibnamefont {An}}, \bibinfo {author} {\bibfnamefont {M.}~\bibnamefont {Xu}}, \bibinfo {author} {\bibfnamefont {A.}~\bibnamefont {Mucchietto}}, \bibinfo {author} {\bibfnamefont {C.}~\bibnamefont {Kim}}, \bibinfo {author} {\bibfnamefont {K.-W.}\ \bibnamefont {Moon}}, \bibinfo {author} {\bibfnamefont {C.}~\bibnamefont {Hwang}},\ and\ \bibinfo {author} {\bibfnamefont {D.}~\bibnamefont {Grundler}},\ }\bibfield  {title} {\bibinfo {title} {Emergent coherent modes in nonlinear magnonic waveguides detected at ultrahigh frequency resolution},\ }\href {https://doi.org/https://doi.org/10.1038/s41467-024-51483-7} {\bibfield  {journal} {\bibinfo  {journal} {Nature Communications}\ }\textbf {\bibinfo {volume} {15}},\ \bibinfo {pages} {7302} (\bibinfo {year} {2024})}\BibitemShut {NoStop}%
\bibitem [{\citenamefont {Rubiola}(2006)}]{rubiola2006measurement}%
  \BibitemOpen
  \bibfield  {author} {\bibinfo {author} {\bibfnamefont {E.}~\bibnamefont {Rubiola}},\ }\bibfield  {title} {\bibinfo {title} {The measurement of am noise of oscillators},\ }in\ \href {https://doi.org/https://doi.org/10.1109/FREQ.2006.275483} {\emph {\bibinfo {booktitle} {2006 IEEE International Frequency Control Symposium and Exposition}}}\ (\bibinfo {organization} {IEEE},\ \bibinfo {year} {2006})\ pp.\ \bibinfo {pages} {750--758}\BibitemShut {NoStop}%
\bibitem [{\citenamefont {Rubiola}(2008)}]{rubiola2008phase}%
  \BibitemOpen
  \bibfield  {author} {\bibinfo {author} {\bibfnamefont {E.}~\bibnamefont {Rubiola}},\ }\href@noop {} {\emph {\bibinfo {title} {Phase noise and frequency stability in oscillators}}}\ (\bibinfo  {publisher} {Cambridge University Press},\ \bibinfo {year} {2008})\BibitemShut {NoStop}%
\bibitem [{\citenamefont {Devitt}\ \emph {et~al.}(2024)\citenamefont {Devitt}, \citenamefont {Wang}, \citenamefont {Tiwari},\ and\ \citenamefont {Bhave}}]{devitt2024edge}%
  \BibitemOpen
  \bibfield  {author} {\bibinfo {author} {\bibfnamefont {C.}~\bibnamefont {Devitt}}, \bibinfo {author} {\bibfnamefont {R.}~\bibnamefont {Wang}}, \bibinfo {author} {\bibfnamefont {S.}~\bibnamefont {Tiwari}},\ and\ \bibinfo {author} {\bibfnamefont {S.~A.}\ \bibnamefont {Bhave}},\ }\bibfield  {title} {\bibinfo {title} {An edge-coupled magnetostatic bandpass filter},\ }\href {https://doi.org/10.1038/s41467-024-51735-6} {\bibfield  {journal} {\bibinfo  {journal} {Nature Communications}\ }\textbf {\bibinfo {volume} {15}},\ \bibinfo {pages} {7764} (\bibinfo {year} {2024})}\BibitemShut {NoStop}%
\bibitem [{\citenamefont {Tikhonov}\ and\ \citenamefont {Litvinenko}(2020)}]{tikhonov2020exchange}%
  \BibitemOpen
  \bibfield  {author} {\bibinfo {author} {\bibfnamefont {V.}~\bibnamefont {Tikhonov}}\ and\ \bibinfo {author} {\bibfnamefont {A.}~\bibnamefont {Litvinenko}},\ }\bibfield  {title} {\bibinfo {title} {Exchange spin waves and their application for diagnostics of the layered structure of epitaxial yig films},\ }\href {https://doi.org/10.1016/j.jmmm.2020.167241} {\bibfield  {journal} {\bibinfo  {journal} {Journal of Magnetism and Magnetic Materials}\ }\textbf {\bibinfo {volume} {515}},\ \bibinfo {pages} {167241} (\bibinfo {year} {2020})}\BibitemShut {NoStop}%
\bibitem [{\citenamefont {Tikhonov}\ \emph {et~al.}(2023)\citenamefont {Tikhonov}, \citenamefont {Lock}, \citenamefont {Ptashenko},\ and\ \citenamefont {Sadovnikov}}]{tikhonov2023excitation}%
  \BibitemOpen
  \bibfield  {author} {\bibinfo {author} {\bibfnamefont {V.}~\bibnamefont {Tikhonov}}, \bibinfo {author} {\bibfnamefont {E.~H.}\ \bibnamefont {Lock}}, \bibinfo {author} {\bibfnamefont {A.}~\bibnamefont {Ptashenko}},\ and\ \bibinfo {author} {\bibfnamefont {A.}~\bibnamefont {Sadovnikov}},\ }\bibfield  {title} {\bibinfo {title} {Excitation of exchange spin waves in the transition layer of the two-layer ferrite-ferrite structure},\ }\href {https://doi.org/10.1016/j.jmmm.2023.171251} {\bibfield  {journal} {\bibinfo  {journal} {Journal of Magnetism and Magnetic Materials}\ }\textbf {\bibinfo {volume} {587}},\ \bibinfo {pages} {171251} (\bibinfo {year} {2023})}\BibitemShut {NoStop}%
\bibitem [{\citenamefont {Tikhonov}\ \emph {et~al.}(2024)\citenamefont {Tikhonov}, \citenamefont {Ptashenko},\ and\ \citenamefont {Sadovnikov}}]{TikhonovPhotonMagPhon2024}%
  \BibitemOpen
  \bibfield  {author} {\bibinfo {author} {\bibfnamefont {V.~V.}\ \bibnamefont {Tikhonov}}, \bibinfo {author} {\bibfnamefont {A.}~\bibnamefont {Ptashenko}},\ and\ \bibinfo {author} {\bibfnamefont {A.}~\bibnamefont {Sadovnikov}},\ }\bibfield  {title} {\bibinfo {title} {Interface mechanism of photon-magnon-phonon conversion in an epitaxial ferrite-dielectric structure},\ }\bibfield  {journal} {\bibinfo  {journal} {Journal of Physics D: Applied Physics}\ }\href {https://doi.org/10.1103/PhysRevB.99.184433} {10.1103/PhysRevB.99.184433} (\bibinfo {year} {2024})\BibitemShut {NoStop}%
\bibitem [{\citenamefont {{Sheshukova}}\ \emph {et~al.}(2014)\citenamefont {{Sheshukova}}, \citenamefont {{Beginin}}, \citenamefont {{Sadovnikov}}, \citenamefont {{Sharaevsky}},\ and\ \citenamefont {{Nikitov}}}]{Sheshukova2014}%
  \BibitemOpen
  \bibfield  {author} {\bibinfo {author} {\bibfnamefont {S.~E.}\ \bibnamefont {{Sheshukova}}}, \bibinfo {author} {\bibfnamefont {E.~N.}\ \bibnamefont {{Beginin}}}, \bibinfo {author} {\bibfnamefont {A.~V.}\ \bibnamefont {{Sadovnikov}}}, \bibinfo {author} {\bibfnamefont {Y.~P.}\ \bibnamefont {{Sharaevsky}}},\ and\ \bibinfo {author} {\bibfnamefont {S.~A.}\ \bibnamefont {{Nikitov}}},\ }\bibfield  {title} {\bibinfo {title} {Multimode propagation of magnetostatic waves in a width-modulated yttrium-iron-garnet waveguide},\ }\href {https://doi.org/10.1109/LMAG.2014.2365431} {\bibfield  {journal} {\bibinfo  {journal} {IEEE Magnetics Letters}\ }\textbf {\bibinfo {volume} {5}},\ \bibinfo {pages} {1} (\bibinfo {year} {2014})}\BibitemShut {NoStop}%
\bibitem [{\citenamefont {Litvinenko}\ \emph {et~al.}(2021{\natexlab{b}})\citenamefont {Litvinenko}, \citenamefont {Sethi}, \citenamefont {Murapaka}, \citenamefont {Jenkins}, \citenamefont {Cros}, \citenamefont {Bortolotti}, \citenamefont {Ferreira}, \citenamefont {Dieny},\ and\ \citenamefont {Ebels}}]{litvinenko2021analog}%
  \BibitemOpen
  \bibfield  {author} {\bibinfo {author} {\bibfnamefont {A.}~\bibnamefont {Litvinenko}}, \bibinfo {author} {\bibfnamefont {P.}~\bibnamefont {Sethi}}, \bibinfo {author} {\bibfnamefont {C.}~\bibnamefont {Murapaka}}, \bibinfo {author} {\bibfnamefont {A.}~\bibnamefont {Jenkins}}, \bibinfo {author} {\bibfnamefont {V.}~\bibnamefont {Cros}}, \bibinfo {author} {\bibfnamefont {P.}~\bibnamefont {Bortolotti}}, \bibinfo {author} {\bibfnamefont {R.}~\bibnamefont {Ferreira}}, \bibinfo {author} {\bibfnamefont {B.}~\bibnamefont {Dieny}},\ and\ \bibinfo {author} {\bibfnamefont {U.}~\bibnamefont {Ebels}},\ }\bibfield  {title} {\bibinfo {title} {Analog and digital phase modulation and signal transmission with spin-torque nano-oscillators},\ }\href {https://doi.org/10.1103/PhysRevApplied.16.024048} {\bibfield  {journal} {\bibinfo  {journal} {Phys. Rev. A}\ }\textbf {\bibinfo {volume} {16}},\ \bibinfo {pages} {024048} (\bibinfo {year} {2021}{\natexlab{b}})}\BibitemShut {NoStop}%
\bibitem [{\citenamefont {Litvinenko}\ \emph {et~al.}(2023)\citenamefont {Litvinenko}, \citenamefont {Kumar}, \citenamefont {Rajabali}, \citenamefont {Awad}, \citenamefont {Khymyn},\ and\ \citenamefont {{\AA}kerman}}]{litvinenko2023phase}%
  \BibitemOpen
  \bibfield  {author} {\bibinfo {author} {\bibfnamefont {A.}~\bibnamefont {Litvinenko}}, \bibinfo {author} {\bibfnamefont {A.}~\bibnamefont {Kumar}}, \bibinfo {author} {\bibfnamefont {M.}~\bibnamefont {Rajabali}}, \bibinfo {author} {\bibfnamefont {A.~A.}\ \bibnamefont {Awad}}, \bibinfo {author} {\bibfnamefont {R.}~\bibnamefont {Khymyn}},\ and\ \bibinfo {author} {\bibfnamefont {J.}~\bibnamefont {{\AA}kerman}},\ }\bibfield  {title} {\bibinfo {title} {Phase noise analysis of mutually synchronized spin hall nano-oscillators},\ }\bibfield  {journal} {\bibinfo  {journal} {Appl. Phys. Lett.}\ }\textbf {\bibinfo {volume} {122}},\ \href {https://doi.org/https://doi.org/10.1063/5.0152381} {https://doi.org/10.1063/5.0152381} (\bibinfo {year} {2023})\BibitemShut {NoStop}%
\bibitem [{\citenamefont {Bianchini}\ \emph {et~al.}(2010)\citenamefont {Bianchini}, \citenamefont {Cornelissen}, \citenamefont {Kim}, \citenamefont {Devolder}, \citenamefont {van Roy}, \citenamefont {Lagae},\ and\ \citenamefont {Chappert}}]{bianchini2010}%
  \BibitemOpen
  \bibfield  {author} {\bibinfo {author} {\bibfnamefont {L.}~\bibnamefont {Bianchini}}, \bibinfo {author} {\bibfnamefont {S.}~\bibnamefont {Cornelissen}}, \bibinfo {author} {\bibfnamefont {J.-V.}\ \bibnamefont {Kim}}, \bibinfo {author} {\bibfnamefont {T.}~\bibnamefont {Devolder}}, \bibinfo {author} {\bibfnamefont {W.}~\bibnamefont {van Roy}}, \bibinfo {author} {\bibfnamefont {L.}~\bibnamefont {Lagae}},\ and\ \bibinfo {author} {\bibfnamefont {C.}~\bibnamefont {Chappert}},\ }\bibfield  {title} {\bibinfo {title} {{Direct experimental measurement of phase-amplitude coupling in spin torque oscillators}},\ }\href {https://doi.org/10.1063/1.3467043} {\bibfield  {journal} {\bibinfo  {journal} {Applied Physics Letters}\ }\textbf {\bibinfo {volume} {97}},\ \bibinfo {pages} {032502} (\bibinfo {year} {2010})}\BibitemShut {NoStop}%
\bibitem [{\citenamefont {Kajfez}\ and\ \citenamefont {Guillon}(1986)}]{kajfez1986dielectric}%
  \BibitemOpen
  \bibfield  {author} {\bibinfo {author} {\bibfnamefont {D.}~\bibnamefont {Kajfez}}\ and\ \bibinfo {author} {\bibfnamefont {P.}~\bibnamefont {Guillon}},\ }\bibfield  {title} {\bibinfo {title} {Dielectric resonators},\ }\bibfield  {journal} {\bibinfo  {journal} {Norwood}\ }\href {https://doi.org/https://doi.org/10.1109/8.250458} {https://doi.org/10.1109/8.250458} (\bibinfo {year} {1986})\BibitemShut {NoStop}%
\bibitem [{\citenamefont {Kajfez}\ and\ \citenamefont {Hwan}(1984)}]{Kajfez1984}%
  \BibitemOpen
  \bibfield  {author} {\bibinfo {author} {\bibfnamefont {D.}~\bibnamefont {Kajfez}}\ and\ \bibinfo {author} {\bibfnamefont {E.~J.}\ \bibnamefont {Hwan}},\ }\bibfield  {title} {\bibinfo {title} {Q-factor measurement with network analyzer},\ }\href {https://doi.org/10.1109/TMTT.1984.1132751} {\bibfield  {journal} {\bibinfo  {journal} {IEEE transactions on microwave theory and techniques}\ }\textbf {\bibinfo {volume} {32}},\ \bibinfo {pages} {666} (\bibinfo {year} {1984})}\BibitemShut {NoStop}%
\end{thebibliography}%

\end{document}